\documentclass[lettersize,journal]{IEEEtran}
\usepackage{amsmath,amsfonts}
\usepackage{algorithmic}
\usepackage{algorithm}
\usepackage{array}
\usepackage[caption=false,font=normalsize,labelfont=sf,textfont=sf]{subfig}
\usepackage{textcomp}
\usepackage{stfloats}
\usepackage{url}
\usepackage{verbatim}
\usepackage{graphicx}
\usepackage{cite}
\usepackage{caption}
\usepackage{tikz}
\usepackage{algorithm}
\usepackage{algorithmic}
\usepackage{multirow}
\usepackage[most]{tcolorbox}
\usepackage{xcolor}
\usepackage{enumitem}
\usepackage{balance}
\usepackage{multicol}
\usepackage{booktabs}

\newcommand{\fram}{SafeCiM}
\newcommand{\blue}[1]{{\color{black}#1}}

\begin{document}

\title{\fram: Investigating Resilience of Hybrid Floating-Point Compute-in-Memory Deep Learning Accelerators}

\author{
    \IEEEauthorblockN{ Swastik Bhattacharya, \textit{Student Member, IEEE}, 
    Sanjay Das, \textit{Student Member, IEEE}, 
    Anand Menon, \textit{Student Member, IEEE}, 
    Shamik Kundu, \textit{Member, IEEE}, 
    Arnab Raha, \textit{Senior Member, IEEE},\\
    Kanad Basu, \textit{Senior Member, IEEE} }

\thanks {S. Bhattacharya, S. Das, A. Menon are with the Department of Electrical and Computer Engineering, University of Texas at Dallas, Richardson, TX 75080 (e-mail: swastikbimal.bhattacharya@utdallas.edu;  sanjay.das@utdallas.edu; anand.menon@utdallas.edu).

S. Kundu, and A. Raha are with the Advanced Architecture Research Group at Intel Corporation, Santa Clara, CA 95054. (e-mail: shamik.kundu@intel.com; arnab.raha@intel.com),

K. Basu is affiliated to the Department of Electrical, Computer and Systems Engineering, Rensselaer Polytechnic Institute, Troy, NY 12180 (e-mail: basuk@rpi.edu).

% The authors declare conflict of interest with all faculties from the University of Texas at Dallas, and Rensselaer Polytechnic Institute.
}}

\maketitle

\begin{abstract}
Deep Neural Networks (DNNs) continue to grow in complexity with the advent of Large Language Models (LLMs), incorporating an increasingly vast number of parameters. The efficient handling of these large volumes of parameters for computations in traditional DNN accelerators is constrained by data transmission bottlenecks. This prompted the exploration of Compute-in-Memory (CiM) architectures that integrate computation units within or near the memory units to address data-transfer bottlenecks. Studies have proposed CiM architectures performing computations in Floating-Point (FP) and Integers (INT). FP computations often yield higher output quality due to their larger representation range and higher precision compared to INT operations, aiding in precision-sensitive applications such as in Generative Artificial Intelligence (GenAI) models, including LLMs and diffusion models, prompting developments and advancements of FP-CiM accelerators. Despite their benefits, the vulnerability of FP-CiM accelerators to hardware faults remains underexplored. This presents a significant challenge, as faults can substantially reduce reliability and can be detrimental in mission critical applications. To address this research gap, we, for the first time, systematically analyze the effect of hardware faults in FP-CiM by introducing bit-flip faults at key computational stages such as digital multipliers, CiM memory cells and digital adder trees. Our experiments with Convolutional Neural Networks (CNNs) such as AlexNet, and state-of-the-art LLMs that include LLaMA-3.2-1B and Qwen-0.3B-Base demonstrate how faults in each CiM computation stage affect the inference accuracy. We show that a single adder fault can bring the model accuracy down to $0\%$ in the case of LLMs. Additionally, based on the insights from the fault analysis, we propose a fault resilient design, \fram{}, that can mitigate fault effect much better than a naive FP-CiM with a pre-alignment stage. For instance,  our \fram{} design for 4096 MAC units can reduce the model accuracy drop by a factor of $49\times$ for a single adder fault compared to a baseline FP-CiM architecture having a pre-alignment stage.
\end{abstract}


\section{Introduction}
\label{sec:introduction}

Generative models such as LLMs are becoming more pervasive in fields such as computer vision, Natural Language Processing (NLP), and autonomous systems~\cite{sarker2021deep}. However, their growing computational footprint demands new hardware paradigms. With the model complexities outpacing the efficiency of traditional von Neumann architectures, minimizing energy-hungry data movement has become a prime concern~\cite{khan2024landscape}. To overcome this limitation, CiM architectures integrate computational logic directly within or near memory arrays, drastically reducing data movement and improving both speed and energy efficiency.

CiM architectures can be of analog or digital variants~\cite{khan2024landscape}. Analog CiM architectures perform computation directly within the analog domain by using Digital-to-Analog Converters (DACs). These convert digital input signals into an analog form. Analog CiMs execute the arithmetic operations in the analog domain, and then convert outputs back into digital format using Analog-to-Digital Converters (ADCs)~\cite{khan2024landscape}. However, analog CiMs face several limitations, such as precision loss during conversion, sensitivity to noise, and process variation \cite{sumbul2023fully}. Digital CiMs, on the other hand, use CMOS logic to perform computations in the digital domain, offering greater reliability and noise resilience \cite{sumbul2023fully, desoli202316, fujiwara20225}. Several digital CiM designs, such as those developed by Samsung \cite{kwon202125, lee2021hardware} and SK Hynix \cite{he2020newton, lee20221ynm}, have shown promising results in memory-centric acceleration for DNN inference.

Modern GenAI models, such as LLMs and diffusion models, typically require FP precision to maintain output quality and generation fidelity particularly critical in safety-sensitive edge-AI applications~\cite{wen2024generative, chen2024low, andreev2022quantization}. Embedding full FP arithmetic in-memory for GenAI models is both area and energy-intensive~\cite{tu2022redcim}. Furthermore, CiM architectures are predominantly optimized for INT computation. To address the area and energy requirement for FP computations on the edge for CiM architectures, accelerators have emerged that decompose exponent and mantissa operations across lightweight digital INT Multiply and Accumulate (MAC) units~\cite{tu2022redcim}. This forms a hybrid architecture, where both INT and FP computations can occur without any significant change in the compute flow~\cite{tu2022redcim}. These architectures have a pipeline to perform FP arithmetic using INT MAC operations, preserving FP accuracy while staying within the constraints of near-memory logic~\cite{tu2022redcim, 10438365, 10716755, 10829928, 10752350, 10268770, 10477933}. 

While FP CiMs present a compelling path forward, their robustness under hardware faults remains largely unexplored. Faults in DNN accelerators can arise from environmental conditions (\textit{e.g.}, radiation), aging, or manufacturing defects~\cite{dixit2021silent, sahoo2016design, zhang2017enabling, borkar2009design, kundu2021toward}. These faults can be transient or permanent. Transient faults, often soft errors, are temporary disruptions~\cite{dixit2021silent}. Permanent faults, such as stuck-at faults, can persist across inferences, leading to repeatable and systematic computation errors~\cite{kundu2021toward}. Unlike control-path faults, which often result in system crashes and are easily detectable, datapath faults silently corrupt numerical computations and propagate across layers. Therefore, it is more challenging to detect and diagnose datapath faults, which can become particularly hazardous in safety-critical systems such as autonomous driving and aerial surveillance, where even minor deviations from the expected model behavior can violate functional safety constraints and result in catastrophic outcomes \cite{10.1145/3638242}.

Prior studies on robustness evaluation of hardware accelerators, such as Google's TPU, have demonstrated how faults can induce NaNs and other silent errors, motivating the advent of robust fault resilience mechanisms like ECC~\cite{he2023understanding, dixit2021silent, rissner2024counting}. Although fault resilience has been extensively studied in conventional DNN hardware accelerators, digital FP-CiM architectures have received limited attention. This creates an urgent need for systematic fault analysis tailored to the unique characteristics of FP-CiM designs.

Fault injection analysis provides a mechanism for evaluating the resilience of DNN accelerators under hardware faults~\cite{9000110}. Existing simulators for CiMs, such as CrossSim \cite{xiao2022crosssim} and NeuroSim \cite{chen2018neurosim}, primarily target analog CiMs and lack support for fault injection. Others, like SIAM \cite{krishnan2021siam} and ZigZag \cite{mei2021zigzag}, focus on digital CiM inference but do not provide the ability to evaluate fault tolerance profiles of FP-CiM. Additionally, low-level hardware simulation is often computationally infeasible for large DNNs. For example, to showcase the fidelity of Register Transfer Level (RTL)-based simulation, the authors of~\cite{9325272} performed fault injection experiments using only a few hundred samples for inference of a small model (LeNet). However, the quadratic growth in computational complexity makes such low-level simulation infeasible for larger models such as LLMs with billions of parameters. Therefore, a scalable fault injection approach for digital FP-CiMs is imperative. Thus, for the first time, we present a thorough analysis into the fault resilience of digital FP-CiM accelerators through scalable fault injection experiments. Drawing the insights from the fault injection experiments, we further propose a fault-aware guideline for designing a robust digital FP-CiM DNN accelerator. Our contributions are:

\begin{itemize}[left=0pt]
\item We perform, for the first time, a comprehensive study of hardware fault resilience in digital FP-CiM accelerators.

\item We assess the sensitivity of each computation stage in the digital FP-CiM, highlighting their vulnerabilities through their effects on DNN inference.

\item We propose, for the first time, a robust FP-CiM design called \textbf{\fram{}} for a case of 4096 MAC units, based on a set of design guidelines derived from the insights of our comprehensive fault resilience study.

\blue{\item Our proposed\textbf{ \fram{}} design shows an improvement of up to 49× in model accuracy due to a single adder fault compared to a baseline pre-alignment design.}
\end{itemize}

Our paper is organized into seven sections. We start with the introduction in Section~\ref{sec:introduction}, followed by a background on previous research and our baseline FP-CiM design in Section~\ref{sec:related_works}. Section~\ref{sec:fault_model} outlines our fault model, and Section~\ref{sec:methodology} describes the fault injection methodology. We analyze the fault injection experiments in Section~\ref{sec:results} and propose the design \textbf{\fram{}} in Section~\ref{sec:design}, concluding with remarks in Section~\ref{sec:conclusion}.

\begin{figure}
    \centering
    \includegraphics[width=0.95\linewidth]{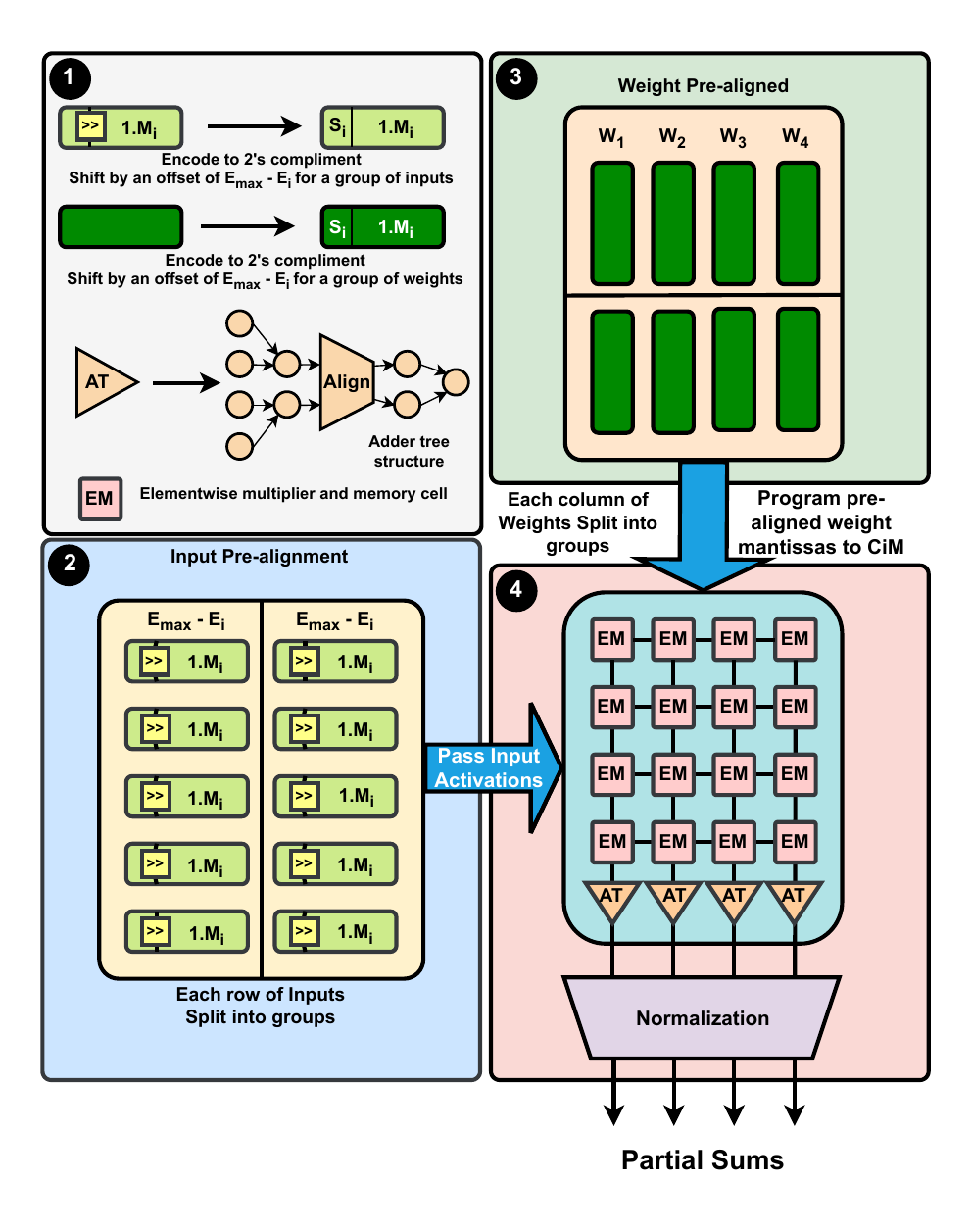}
    \caption{Floating-point CiM architecture.}
    \label{fig:fpcim}
\end{figure}

\section{Background and Related Works}
\label{sec:related_works}

This section provides a background into the recent studies that have been performed towards the development of different CiM architectures. We build on these to outline the base digital FP-CiM architecture that is used in our fault injection experiments. We also provide a background into the past research on impact of faults on different compute devices.

\subsection{FP-CiM Architectures}
\label{subsec:fpcim_architecture}
CiM architectures provide a solution to the memory bottleneck in von Neumann architectures by embedding MACs in memory. Early analog RRAM/PCM crossbars exploited Ohm's and Kirchhoff's laws for in-memory MACs \cite{10546882,hu2016dot}. However, they suffer from noise, device variability, and limited precision that hinder DNN and LLM inference \cite{sumbul2023fully}. Digital CiM on CMOS SRAM or gain-cell arrays now affords greater reliability and native BF16/FP16 support \cite{10268770,10829928,10716755}. These designs translate floating-point workloads into integer hardware: mantissas are exponent-aligned, multiplied via in-memory INT MACs, partial sums accumulated, and results normalized. Architectures such as ReDCIM \cite{tu2022redcim}, FP-IMC \cite{10268770}, and dual-mode gain-cell arrays \cite{10716755} employ Booth-encoded signed multiplication with direct INT accumulation. They consist of unified INT/FP pipelines performing calculations in both modes seamlessly within each cell \cite{10829928}. A central trade-off is the position of mantissa alignment in the FP-CiM computation steps. Pre-alignment (\textit{e.g.}, in ReDCIM) shifts mantissas and resolves signs before the array, preserving simple INT accumulation but adding front-end logic. Post-alignment (\textit{e.g.}, in FP-IMC) defers alignment until after multiplication, reducing preprocessing at the cost of extra hardware.

For our fault injection experiments to assess hardware resilience, we use a pre-alignment design as a baseline. Fig.~\ref{fig:fpcim} illustrates the schematic of the architecture. Part \tikz[baseline=-0.5ex]\node[circle,fill=black,text=white,inner sep=0.5mm]{1}; of the figure shows the symbols and their definitions used for the baseline FP-CiM design schematic. Part \tikz[baseline=-0.5ex]\node[circle,fill=black,text=white,inner sep=0.5mm]{2}; of the figure describes the input alignment of this pre-alignment design baseline. Each of the input rows is partitioned into groups of size $g$ elements. Within each group, mantissas shift to the group's maximum exponent, and are converted to two's-complement via the inputs' sign bit. Part \tikz[baseline=-0.5ex]\node[circle,fill=black,text=white,inner sep=0.5mm]{3}; of the figure shows the pre-alignment of weights, similar to that for inputs, where each column is split to different groups for alignment. Part \tikz[baseline=-0.5ex]\node[circle,fill=black,text=white,inner sep=0.5mm]{4}; shows the CiM computation pipeline that maps scalar multiply–add onto near-memory compute blocks. 

In the above design, an \textbf{IC $\times$ OC} crossbar array holds the aligned weights, and co-located multipliers compute integer products. \textbf{IC} stands for the Input Channels (number of rows in the crossbar array), and \textbf{OC} stands for Output Channels (number of columns). For each output column of the crossbar, the \textbf{IC} products feed an adder tree, with summations upto $\log_2 g$ levels of the tree. Intermediate sums after these adder levels are tied to a different local exponent. These sums are re-aligned to the column's global maximum exponent. Then, the aligned sums are added over the remaining of the total $\log_2{\textbf{IC}}$ adder levels. Finally, the sign is separated, the mantissa normalized to restore a leading `1', and the exponent adjusted to yield the final FP partial sum \cite{coonen1980implementation}. We adopt this FP-CiM architecture as the baseline for fault injection experiments. 

\blue{\subsection{Prior Works on Silent Data Corruption due to Faults}
\label{subsec:fault_inj_works}
Prior resilience studies inform our work; however, they do not address the specific challenges of CiM. Veritas~\cite{chatzopoulos2025veritas} quantifies Silent Data Corruption (SDC) trends from arithmetic-datapath faults in conventional CPU pipelines, yet it omits the distinctive crossbar operation sequence and alignment stages of CiM accelerators. Shoestring \cite{feng2010shoestring} reduces soft-error rate by duplicating selectively vulnerable instructions, but it neither probes arithmetic-datapath faults nor considers CiM architectures. Accelerated SFI frameworks such as SASSIFI~\cite{hari2017sassifi}, MeRLiN~\cite{kaliorakis2017merlin}, and AVGI~\cite{papadimitriou2023avgi} inject errors into general-purpose registers, memory, and predicate registers of CPU/GPU cores, again leaving CiM pipelines and datapath faults unexplored. In contrast, our study performs bit-level fault injection across every stage of the FP-CiM pipeline—including the arithmetic datapath. Its scalable methodology supports billion-parameter LLMs, overcoming the prohibitive runtime of RTL-level simulations. 
}

\section{Fault Model}
\label{sec:fault_model}

\blue{In this work, we focus on datapath faults in the digital FP-CiM accelerator that can occur in all the computation stages such as exponent comparison, offset calculation, mantissa alignment, multiplication, addition, global alignment and normalization discussed in Section~\ref{subsec:fpcim_architecture}. Unlike control-path faults that trigger detectable crashes, they silently corrupt numerical computations and degrade DNN inference accuracy. These data corruptions can lead to catastrophic outcomes in safety-critical systems (\textit{e.g.}, misinterpreted speed-limit signs in autonomous vehicles)~\cite{kundu2021toward}. Hardware faults are either transient or permanent (refer Section~\ref{sec:introduction}). We target permanent faults due to their persistence, which can lead to consistent degradation in system performance, such as DNN accuracy when running a task on faulty hardware. This effect cannot be masked (unlike for transient faults) and would require costly offline testing for mitigation not feasible in mission-critical applications~\cite{zhang2012runtime}.

To assess the effects of permanent faults in FP-CiM, we inject bit-flips in the CiM computation datapath at fixed locations active throughout each DNN inference~\cite{staudigl2023fault,highresilientAES}. This bit-flip fault model will always induce errors regardless of the original bit state, unlike stuck-at fault model~\cite{maddah2012data}. This makes bit-flip fault model more stringent than stuck-at fault model for a worst-case impact analysis~\cite{chang2018hamartia,richter2021fiver}. Past research has also shown that voltage fluctuations induce faults manifesting as bit-flips in DNN hardware with bit error rates upto $1\%$~\cite{sun2024neuralfuse}. These motivate us to use a bit-flip fault model for our experiments.

Our fault model focuses in single-bit faults. It has been demonstrated as a realistic fault model, with NVIDIA reporting that $78\%$ faults under a particle-strike model manifest themselves as single-bit fault~\cite{chang2018hamartia}. Multi-bit faults are comparatively difficult to model in comparison to single-bit faults since their effect is primarily dependent on the distribution of the bits values~\cite{chang2018hamartia}. Therefore, we use a single-bit fault model for our experiments.

We select the locations for fault injection assuming a uniform likelihood for a fault to occur in a computation unit of the FP-CiM compute pipeline. Although Weibull distributions are widely used to model time-to-failure of hardware units (\textit{e.g.}, manufacturing defects or wear-out)~\cite{toshiba2017reliability, minitab2025weibull}, they are not typically employed to decide which computation units will have a fault at a single point in time. Instead, most fault-injection studies adopt uniform random sampling of compute units as a baseline~\cite{colucci2023isimdl, parikh2013udirec, moradi2023failure, benso2011art, white2021establishing, kaja2024statistical}. 
}

\section{Methodology}
\label{sec:methodology}

In this section, introduce our fault injection framework, \textbf{FaultCiM}, for executing our fault injection experiments.

\subsection{FaultCiM Framework Overview}
\label{subsec:framework_overview}
\begin{figure}[t]
    \centering
    \includegraphics[width=0.95\linewidth]{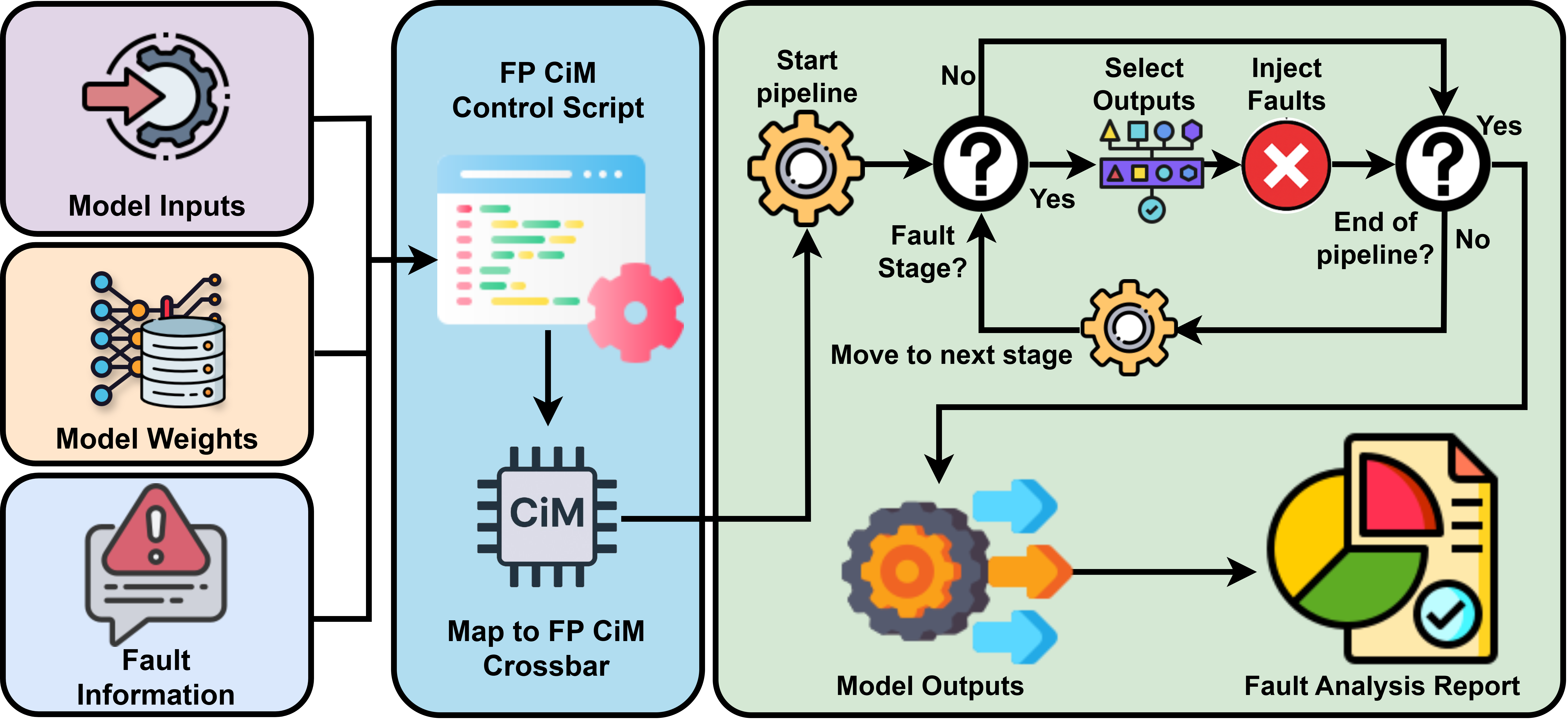}
    \caption{FaultCiM overview.}
    \label{fig:framework_overview}
\end{figure}

To analyze how hardware faults in digital FP-CiM datapaths affect DNN inference, we use \textbf{FaultCiM}. FaultCiM is our fault-injection framework that injects faults at different computation stages described in Section~\ref{subsec:fpcim_architecture}, and measures the resilience of digital FP-CiM accelerators to hardware faults. To use FaultCiM, as illustrated in Fig.~\ref{fig:framework_overview}, the user specifies five parameters: (1) model parameters, (2) input dataset, (3) the pipeline stage for fault injection, (4) number of faults to be injected in terms of percentage of total computations in the stage, and (5) the bit position for fault injection via bit-flip. Model weights and inputs are mapped onto the CiM crossbar, which then executes the six-stage compute flow—\textbf{pre-alignment} of inputs and weights, programming aligned weight mantissas to \textbf{CiM memory} cells, integer multiplication of programmed weight mantissas and aligned inputs by \textbf{CiM multipliers}, summation of mantissa products by \textbf{CiM Adder tree}, \textbf{global alignment} of mantissa sums, and \textbf{normalization} of the mantissa sums. At each stage, FaultCiM interjects the computations and injects bit-flip faults at specified locations. The corrupted outputs then propagate downstream the FP-CiM computation pipeline. Since the framework is modular, it can evaluate fault effects across individual architectural blocks and diverse workloads.

\subsection{Fault Injection}
\label{subsec:fault_injection}
\begin{figure*}[!tp]
\centering
\includegraphics[width=0.95\linewidth]{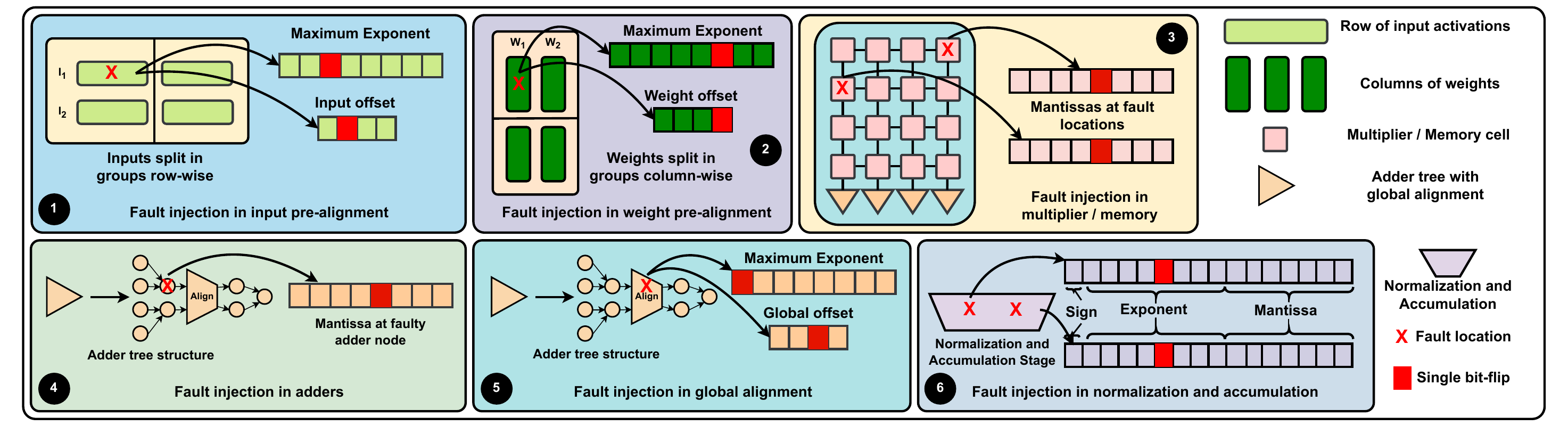}
\caption{Bit-flip fault injection locations in FP CiM datapath.}
\label{fig:fault_modeling}
\end{figure*}

Single bit-flips, used to model faults as per Section~\ref{sec:fault_model}, are introduced at the $i^{th}$ bit of a scalar $x$ using an appropriate bit mask. FaultCiM applies this fault model across different computation stages of FP-CiM. Fig.~\ref{fig:fault_modeling} shows where in the the various computation stages FaulCiM injects bit-flip faults. In the \textbf{Pre-Alignment} stage, faults are injected in the maximum exponents or in the offsets calculated using these exponents (Fig.~\ref{fig:fault_modeling}, part \tikz[baseline=-0.5ex]\node[circle,fill=black,text=white,inner sep=0.5mm]{1}; for inputs, and part \tikz[baseline=-0.5ex]\node[circle,fill=black,text=white,inner sep=0.5mm]{2}; for weights). Faults here may misidentify the group-wise maximum exponent or corrupt offset computation, causing mantissa misalignment and degraded precision. In the \textbf{CiM Memory cells}, faults are injected into the datapath for programmed weight matissas (Fig.~\ref{fig:fault_modeling}, part \tikz[baseline=-0.5ex]\node[circle,fill=black,text=white,inner sep=0.5mm]{3};). These faults affect aligned weights reused across crossbar rows, providing corrupted weights to the multipliers for matrix multiplication. In the \textbf{CiM Multipliers}, faults are injected into the multiplier outputs (see Fig.~\ref{fig:fault_modeling}, part \tikz[baseline=-0.5ex]\node[circle,fill=black,text=white,inner sep=0.5mm]{3};). These faults distort individual mantissa products. In the \textbf{CiM Adder} stage, faults are injected into the adder outputs at different levels of the adder tree (Fig.~\ref{fig:fault_modeling}, part \tikz[baseline=-0.5ex]\node[circle,fill=black,text=white,inner sep=0.5mm]{4};). The adder tree, which sums the multiplier products, amplifies faults through successive levels. Faults in the \textbf{Global Alignment} stage are injected either in the exponents of the intermediate adder sums or in the offsets calculated using these exponents (Fig.~\ref{fig:fault_modeling}, part \tikz[baseline=-0.5ex]\node[circle,fill=black,text=white,inner sep=0.5mm]{5};). Faults in exponent addition or offset logic mis-scale entire columns, affecting the adder tree outputs. Finally, faults in the \textbf{Normalization} stage are injected in any of the bits of the resultant FP output (Fig.~\ref{fig:fault_modeling}, part \tikz[baseline=-0.5ex]\node[circle,fill=black,text=white,inner sep=0.5mm]{6};). Bit-flip faults can invert the sign or produce Infinite/Not-a-Number (INF/NaN) outputs, which may not be corrected downstream. Using fault injection, we conduct a comprehensive investigation of these fault manifestations in digital FP-CiM architectures for the first time, aiming to assess FP-CiM resilience and identify design guidelines that yield robust, fault-tolerant CiM accelerators.

\section{Experimental Results}
\label{sec:results}
This section evaluates the resilience of digital FP-CiM modules to bit-flip faults injected at key computational stages outlined in Section~\ref{subsec:fault_injection}. We systematically assess how faults in FP-CiM affect inference accuracy. 

\subsection{Experiment Setup}
\label{subsec:exp_setup}
\begin{table}[h]
    \centering
    \caption{CiM Crossbar Architecture Specifications}
    \label{tab:cim_arch}
    \begin{tabular}{ll}
        \toprule
        \textbf{Parameter} & \textbf{Specification} \\
        \midrule
        Crossbar Size (\textbf{IC} $\times$ \textbf{OC})    & 128 rows $\times$ 32 columns \\
        Pre-alignment Group Size ($g$) & 16 elements \\
        Floating-Point Format & BFLOAT16 \\
        Mantissa Bits  & 7 bits (padded to 12 bits) \\
        Mantissa Representation & 13-bit integers \\
        Mantissa Multiplication & 26-bit elementwise products \\
        Adder Tree Stages  & 4 initial, 3 additional \\
        Final Mantissa Width & 33 bits \\
        \bottomrule
    \end{tabular}
\end{table}

\subsubsection{\textbf{CiM Architecture}} 

Table~\ref{tab:cim_arch} summarizes the FP-CiM architecture: a $128\times32$ CiM crossbar (in an \textbf{IC $\times$ OC} stencil as per Section~\ref{subsec:fpcim_architecture}) supporting BFLOAT16, chosen to accommodate extended LLM sequences and keys~\cite{laguna2022hardware}. BFLOAT16 encodes values as $(-1)^S \times 2^{E-127} \times 1.M$ with one sign bit $S$ (bit 15), eight exponent bits $E$ (bits 14–7), and seven mantissa bits $M$ (bits 6–0). The mantissa is expanded to 12 bits (by prepending a `1' and appending four zeros) and converted to a 13-bit two’s-complement representation. Pre-alignment is performed in groups of 16, then the 13-bit mantissas are multiplied to yield 26-bit products. Each column generates 128 products, which are then accumulated via a 7-stage ($\log_2 128$) binary adder tree, with a global alignment at stage 4 decided by the pre-alignment group size ($\log_2 16$). Finally, the 33-bit adder sum is normalized, where first the mantissa is converted to its 32-bit unsigned form after extracting the sign bit, followed by bit shifts so that it is in the final $1.M$ form as per BFLOAT16 definition. The direction and number of shifts then changes the exponent values accordingly, to result in a final 8-bit exponent. 7 MSB bits are extracted from $M$ of the 32-bit $1.M$ to form the mantissa bits of the BFLOAT16 resultant. These bits, with the final exponent and extracted sign bit are merged to form the BFLOAT16 output.

\subsubsection{\textbf{DNN models and Datasets with Baseline}} \label{subsubsec:dataset} 
\blue{Fault injection experiments were carried out on four DNN inference workloads mapped to our FP-CiM architecture. AlexNet fine-tuned on CIFAR-100~\cite{krizhevsky2009learning} achieves $70.8\%$ basline accuracy, and BERT-Base (denoted as BERT) on the Emotion tweets corpus reaches $93.75\%$~\cite{saravia2018carer}. On 120 prompts from MMLU, LLaMA-3.2-1B scores $55\%$ \cite{grattafiori2024llama, hendrycks2021measuring} and Qwen-0.6B-Base $53.5\%$. Our experiments use 120 MMLU prompts randomly selected from across all the tasks. We use 120 prompts, as fault injection experiments on these models are computationally prohibitive. These benchmarks spanning over image classification, sentence-level text classification, and autoregressive text generation tasks, cover compute-bound (AlexNet, BERT) and memory-bound (LlaMA-3.2-1B and Qwen-0.3B-Base) inferences, representing a variety of edge-scale workloads. All experiments use PyTorch on an NVIDIA A100-SXM4 GPU.}

\subsubsection{\textbf{Performance Metric for Fault Experiments}} \label{subsubsec:metrics} 
\blue{We measure model accuracy under faults as the fraction of correctly predicted samples among the total samples. Single-bit faults are injected at various computation stages. The amount of injected faults is quantified by the percentage of faulty outputs among total outputs for a computation stage.  For each experiment, the fault locations are initially identified, and the corresponding compute units are affected and maintained throughout the experiment. For instance, injecting faults in $0.1\%$ of the multipliers in a $128\times32$ FP-CiM crossbar means outputs of four pre-determined multipliers are corrupted as discussed in Section~\ref{subsec:fault_injection}. The fraction of faulty computations is converted to Bit-Error Rate (BER) using:

\[
\text{BER} = \frac{S_F}{S_T \times N_B},
\]

where $S_F$ and $S_T$ are the counts of faulty and total outputs from an FP-CiM stage, and $N_B$ is the output bit-width for that stage.
}

\subsection{Experimental Analysis}
\label{subsec:exp_analysis}
\subsubsection{\textbf{Pre-Alignment Fault in Input}}
\label{subsubsec:pre_align_input}

\begin{figure}[t]
    \centering
    \captionsetup{justification=centering}
    \includegraphics[width=0.95\linewidth]{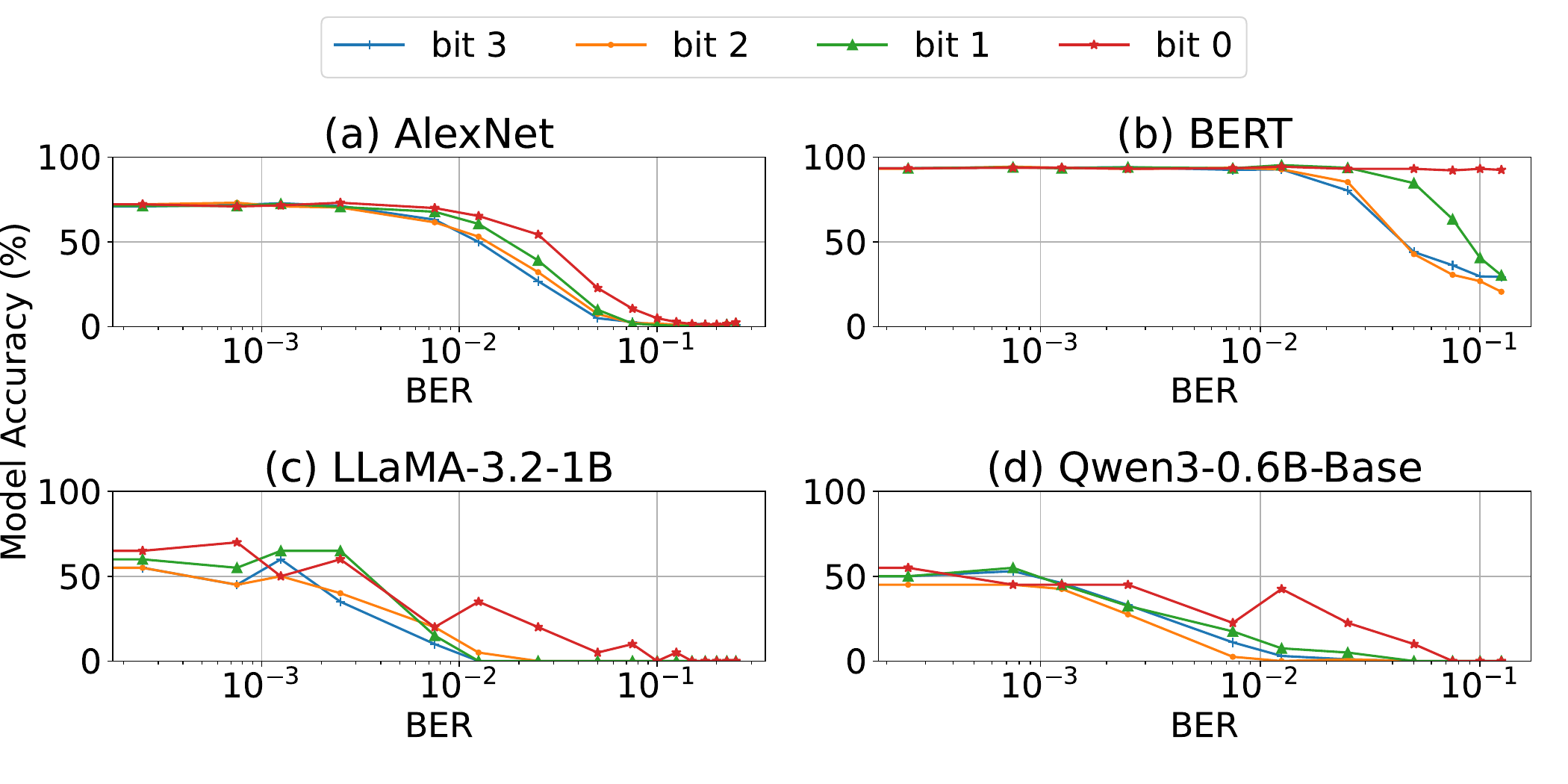}
    \caption{Accuracy due to faults in pre-alignment input offsets.}
    \label{fig:pre_align_input_offsets}
\end{figure}

Fault injection is performed at the Pre-Alignment stage for inputs targeting offset values and maximum exponents among the inputs. 

\paragraph{\textbf{Faults in Offsets}}\noindent Faults in offset computation during input pre‐alignment misalign mantissas before multiplication. We inject bit‐flip faults, affecting a percentage of computations, at different bit positions of the 4‐bit offsets ($N_B=4$). Figs.~\ref{fig:pre_align_input_offsets}(a–d) show accuracy degradation for AlexNet, BERT, LLaMA-3.2-1B, and Qwen3-0.6B-Base due to fault injection. Accuracy declines with increase in BER, with MSB (bit 3) faults causing the steepest drop due to mantissa misalignment by $2^3$ bits. LLaMA-3.2-1B reaches $0\%$ accuracy at $1.25\times10^{-2}$ BER; AlexNet accuracy degrades to $2.81\%$ at $7.5\times10^{-2}$ BER; BERT shows the least degradation. Faults in bits 2, 1, and 0 (LSB) have progressively less impact on model accuracy. 

\begin{figure}[tp]
    \centering
    \captionsetup{justification=centering}
    \includegraphics[width=0.95\linewidth]{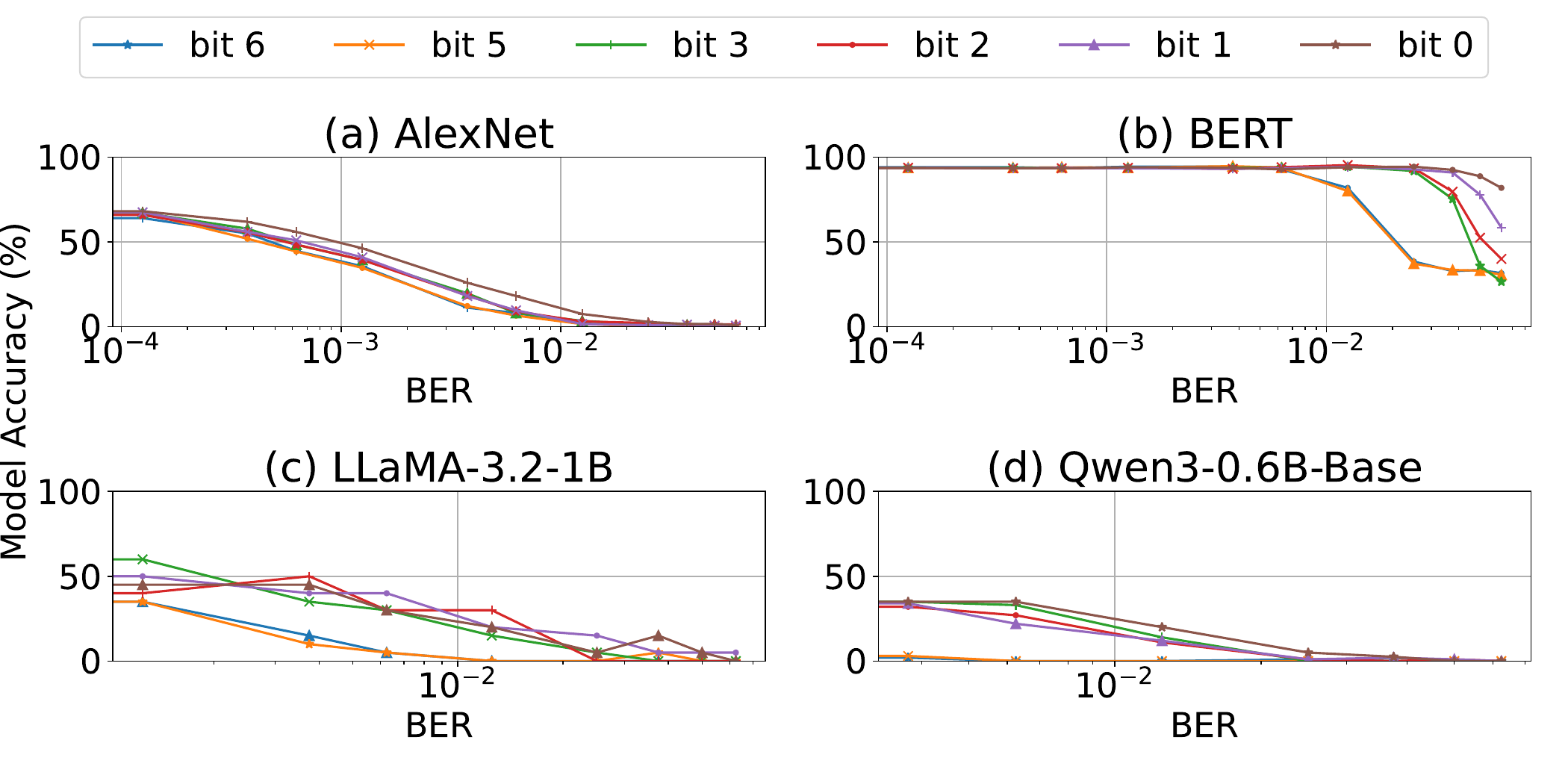}
    \caption{Accuracy due to faults in  pre-alignment input exponents.}
    \label{fig:pre_align_input_exp}
\end{figure}

\paragraph{\textbf{Faults in Exponents}}\noindent Faults in maximum exponent ($N_B=8$) cause exponential scaling errors with mantissa misalignment. Flips in MSB–1 (bit 6) induce an error of factor $2^{\pm64}$. Faults in MSB can lead to NaNs resulting in compute failure, thus we inject faults into bits 6 to 0. Figs.~\ref{fig:pre_align_input_exp}(a–d) show accuracy degradation due to these faults. The accuracy drop is the most for MSB-1 fault, decreasing for bits near the LSB. While AlexNet and LLaMA-3.2-1B remain sensitive to LSB faults, BERT is more resilient due to sparse inputs.

\begin{tcolorbox}[colback=green!5!white,colframe=green!75!black]
\textbf{Observation 1:}  
Faults near MSB, either in the 4-bit offsets or 8-bit exponents, induce catastrophic, layer-wise error buildup and accuracy collapse. Exponent faults are more sensitive than offset faults.
\end{tcolorbox}

\subsubsection{\textbf{Pre-Alignment Fault in Weights}}
\label{subsubsec:pre_align_weights}

Fault injection is performed at the Pre-Alignment of weights to target: (1) offset values; and (2) maximum exponents among the weights. 

\begin{figure}[t]
    \centering
    \captionsetup{justification=centering}

    \includegraphics[width=0.95\linewidth]{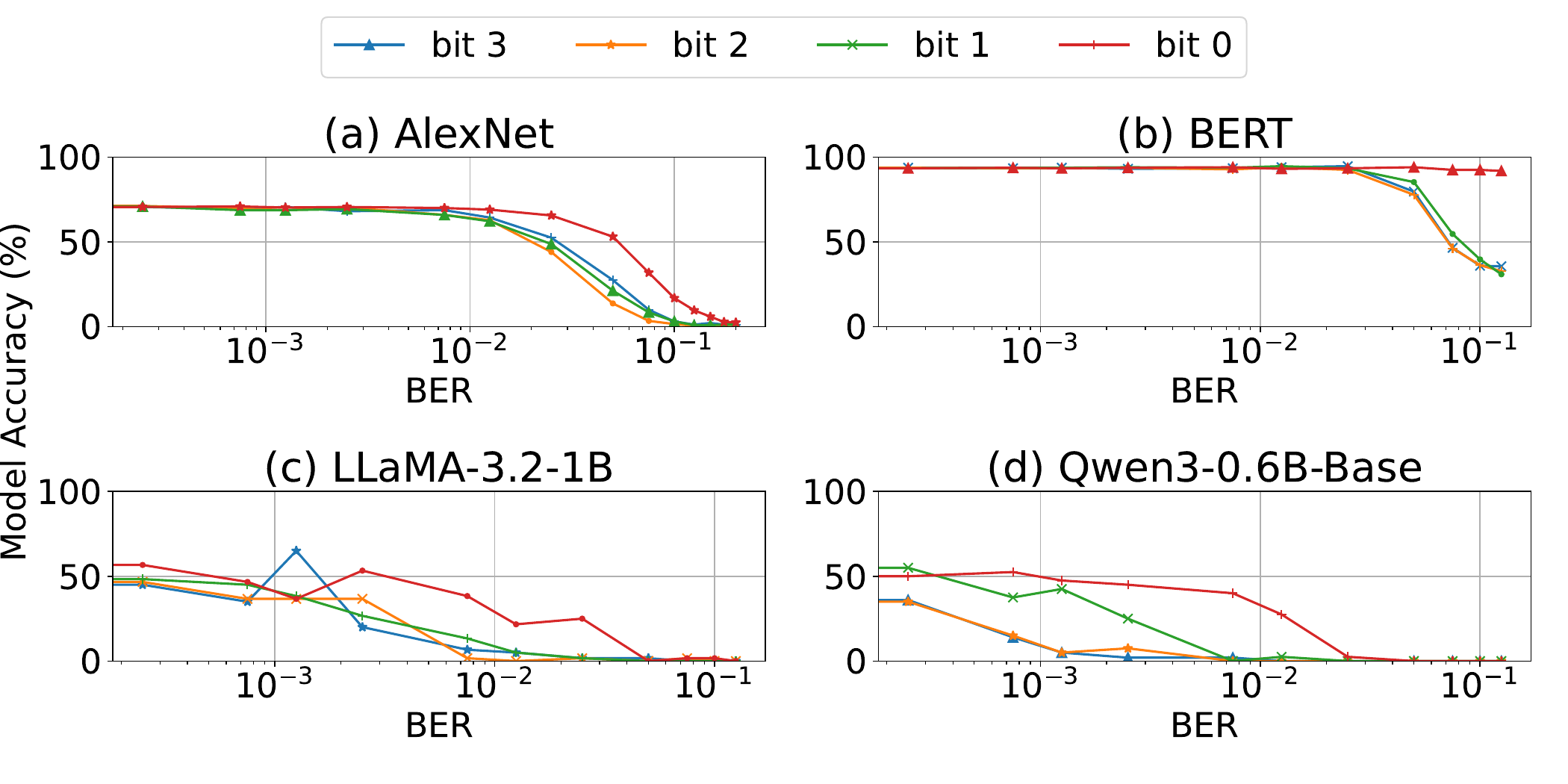}
    \caption{Accuracy due to faults in pre-alignment weight offsets.}
    \label{fig:pre_align_weight_offset}
\end{figure}

\paragraph{\textbf{Faults in Offsets}}\noindent Faults in weight pre‐alignment offsets misalign mantissas in the FP-CiM pipeline. For 4-bit offsets ($N_{B}=4$), bit-flip faults at increasing BER degrade accuracy across AlexNet, BERT, LLaMA-3.2-1B and Qwen3-0.6B-Base as shown in Figs.~\ref{fig:pre_align_weight_offset}(a–d). MSB (bit 3) faults cause the sharpest accuracy drop, followed by bits 2 to 0. Qwen3-0.6B-Base accuracy falls to $5\%$ at BER of $7.5\times10^{-3}$ and $0\%$ at BER of $2.5\times10^{-2}$. LLaMA-3.2-1B reaches $20\%$ accuracy at a smaller BER of $2.5\times10^{-3}$ due to error accumulations across larger layers and deeper model architecture. %Faults have comparatively less effect on BERT. 
LSB faults have a negligible effect and MSB faults lower the accuracy of BERT to $46.56\%$ at a greater BER of $7.5\times10^{-2}$, whereas AlexNet drops to $3.44\%$ at the same BER. 

\begin{figure}[t]
    \centering
    \captionsetup{justification=centering}

    \includegraphics[width=0.95\linewidth]{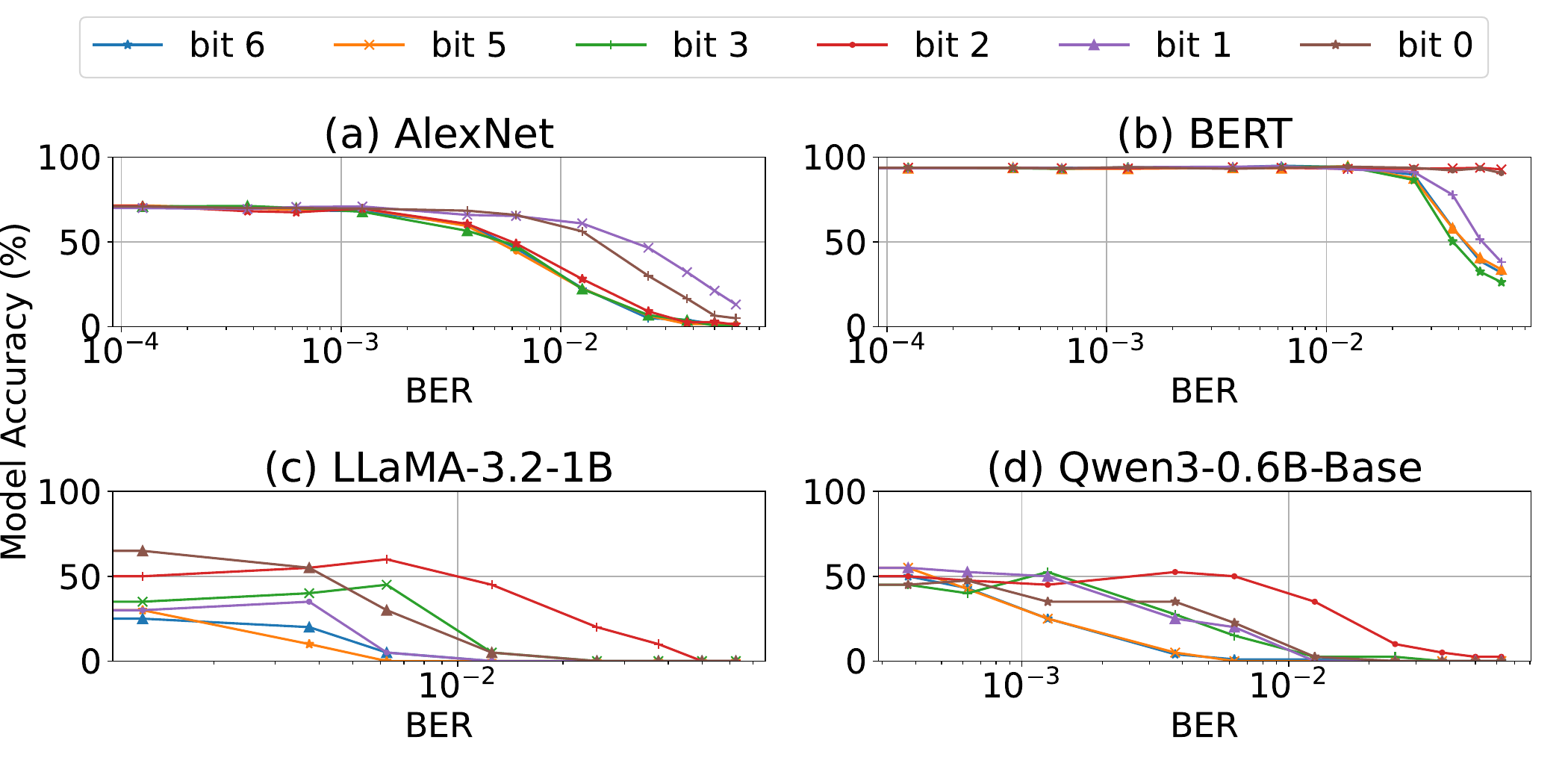}
    \caption{Accuracy due to faults in pre-alignment weight exponents.}
    \label{fig:pre_align_weight_exp}
\end{figure}

\paragraph{\textbf{Faults in Exponents}}\noindent  Faults in weight pre‐alignment exponents introduce multiplicative errors along with severe mantissa misalignment in the FP-CiM pipeline. For 8-bit exponents ($N_{B}=8$), faults injected into bits 6 (MSB–1), 5, 3, 2, 1, and 0 (excluding the MSB to avoid NaNs) yield accuracy degradation as seen in Figs.~\ref{fig:pre_align_weight_exp}(a–d). AlexNet is most sensitive to faults in bits 6 to 2 (bits 1 and 0 have less steep decline); BERT to bits 6 to 3 followed by bit 1 (bits 2 and 0 have minimal effect). LLaMA-3.2-1B is most sensitive to faults in bits 6 and 5, followed by bit 1 (bit 2 has the least effect). These non-uniform trends among the bits are due to exponent faults, which both mis-scale mantissas and shift some of their bits left or right. These opposing effects can compound or diminish the errors due to faults. However, there is a general trend of MSB being more critical than LSB.

\begin{tcolorbox}[colback=green!5!white, colframe=green!75!black]
\textbf{Observation 2:}  
MSB faults in pre-alignment weight offsets and exponents cause mantissa misalignment and scaling errors, severely impacting model accuracy. The accuracy decline varies by model.
\end{tcolorbox}

\subsubsection{\textbf{CiM Memory Faults}}
\label{subsubsec:cim_memory}

\begin{figure}[t]
    \centering
    \captionsetup{justification=centering}

    \includegraphics[width=0.95\linewidth]{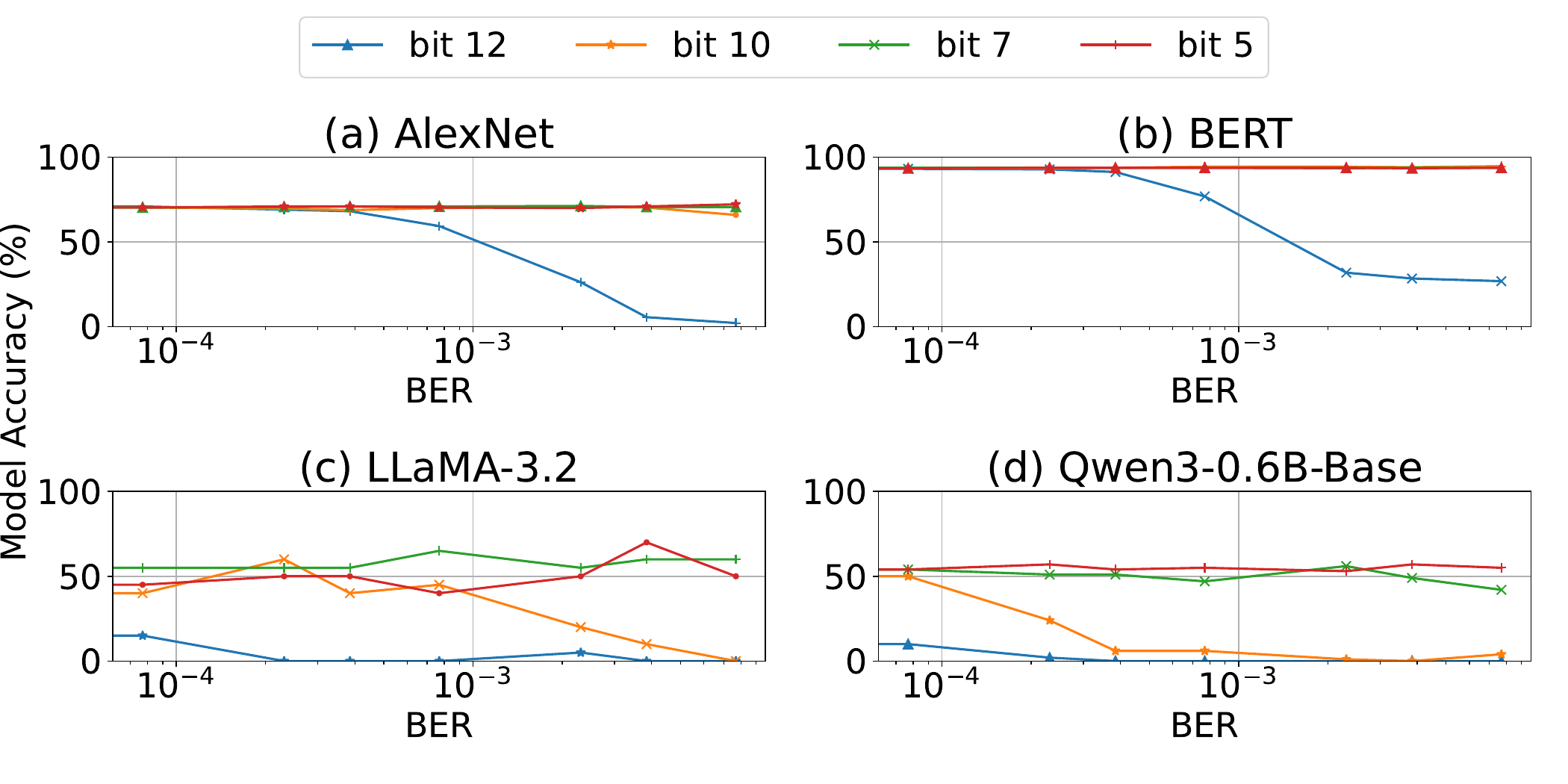}
    \caption{Accuracy after fault injection in CiM memory cells.}
    \label{fig:cim_memory}
\end{figure}

Once weight mantissas are loaded into FP-CiM memory, they are reused across inputs. To assess the effect of faults in CiM Memory cells, single-bit faults are injected into bits 12 (MSB), 10, 7, and 5 of the 13-bit mantissas ($N_B = 13$). Figs.~\ref{fig:cim_memory}(a), (b), (c), and (d) show the accuracy degradation due to faults for AlexNet, BERT, LLaMA-3.2-1B, and Qwen3-0.6B-Base, respectively. Accuracy declines consistently with increasing BER, with severity depending on bit position. LSB faults (e.g., bit 5) have minimal effect, while MSB faults (bit 12) induce sharp degradation with respect to BER, dropping LLaMA-3.2-1B accuracy to $0\%$ at just $7.69\times 10^{-4}$ BER. Similarly, the accuracy of Qwen3-0.6B-Base drops to $0\%$ at just  $2.31\times 10^{-4}$ BER. LLaMA-3.2-1B and Qwen3-0.6B-Base completely fail (accuracy down to $0\%$) due to bit 10 faults affecting them at a BER of $7.69\times 10^{-3}$ and $2.31\times 10^{-2}$ respectively. These faults generally have minimal impact on AlexNet and BERT. This indicates that the deep model is more susceptible to persistent errors that accumulate over the course of DNN inference.

\begin{tcolorbox}[colback=green!5!white, colframe=green!75!black] \textbf{Observation 3:} Single-bit flips near the MSB of the programmed weight mantissa corrupt subsequent CiM operations for BER in the order of $10^{-4} - 10^{-3}$. Faults in LSBs have a negligible impact on CiM integrity. 
\end{tcolorbox}

\subsubsection{\textbf{CiM Multiplier Faults}}
\label{subsubsec:cim_multipliers}

\begin{figure}[b]
    \centering
    \includegraphics[width=0.95\linewidth]{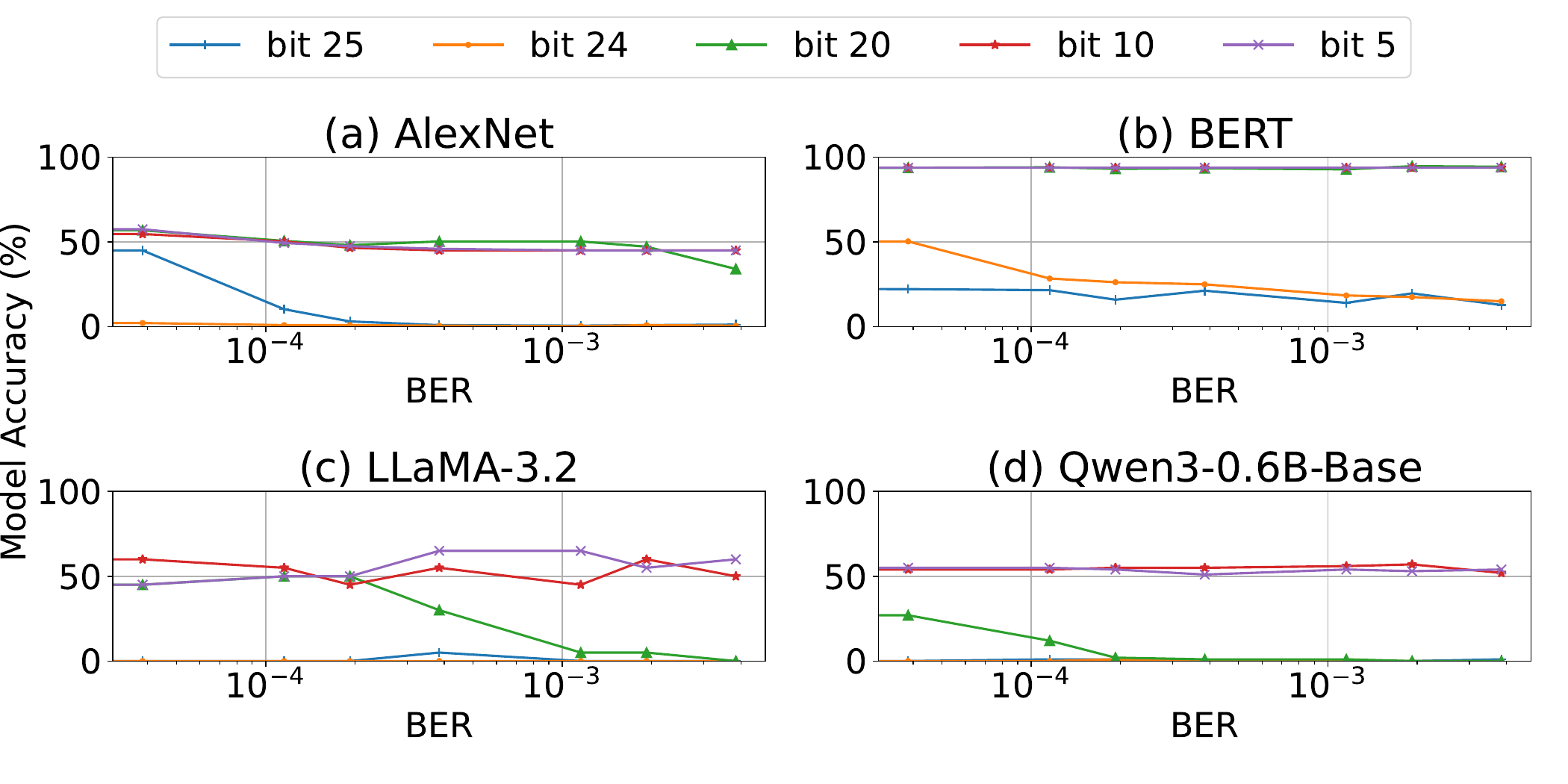}
    \caption{Accuracy after fault injection in CiM multipliers.}
    \label{fig:cim_multipliers}
\end{figure}

Faults in elementwise multiplications directly corrupt the input–weight mantissa products. These products are added to form the mantissa sums for the DNN output. To assess the impact of faults in these products, single-bit faults are injected into select positions of the 26-bit integer multiplier output ($N_B = 26$) —specifically bits 25 (MSB), 24, 20, 10, and 5—to cover high, mid, and low significance bits. Figs.~\ref{fig:cim_multipliers}(a), (b), (c), and (d) show the accuracy variation due to faults for AlexNet, BERT, LLaMA-3.2-1B, and Qwen3-0.6B-Base, respectively. Across all models, accuracy drops progressively with increasing BER, but MSB faults induce the steepest degradation. For AlexNet, faults in bits 25 and 24 reduce accuracy to $3.13\%$ at just $1.29\times 10^{-4}$ BER. Model accuracy of BERT reaches $15.94\%$ at $1.29\times 10^{-4}$ BER, while the accuracy of LLaMA-3.2-1B and Qwen3-0.6B-Base collapse to $0\%$ at only $3.85\times 10^{-5}$ BER. While less severe than in the cases of AlexNet and BERT, faults in bits between MSB and LSB such as bit 20 also reduce accuracy (\textit{e.g.}, $5\%$ accuracy at $1.15\times 10^{-3}$ BER in LLaMA-3.2-1B and $1\%$ at $1.15\times 10^{-3}$ BER in Qwen3-0.6B-Base). On the other hand, because of their little numerical importance, faults in bits 10 to the LSB have minimal effect. 

\begin{tcolorbox}[colback=green!5!white, colframe=green!75!black] 
\textbf{Observation 4:} MSB faults in CiM multiplier outputs (\textit{e.g.}, bits 25–24) show multipliers are highly vulnerable. Faults occurring at BERs of $10^{-5}$–$10^{-4}$ degrade model accuracy more than those in programmed weight mantissas at a similar BER. 
\end{tcolorbox}

\subsubsection{\textbf{CiM Adder Faults}}
\label{subsubsec:cim_adders}

\begin{figure}[t]
    \centering
    \includegraphics[width=0.95\linewidth]{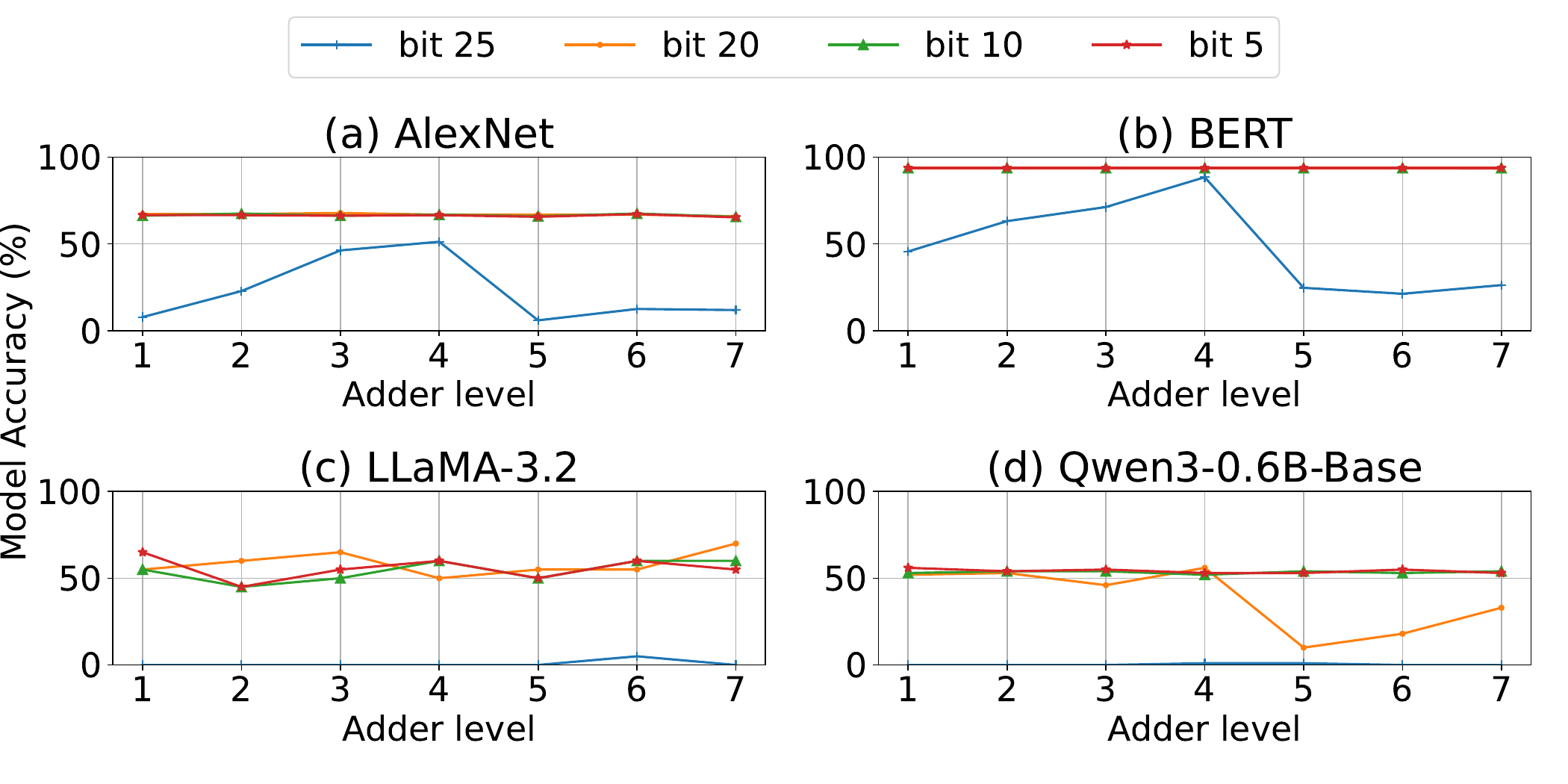}
    \caption{Accuracy after fault injection in CiM adders.}
    \label{fig:cim_adders}
\end{figure}

The adder trees aggregate element-wise products into mantissa partial sums. To assess the device vulnerability to faults in these adder trees, we flipped a single output bit per tree level. Faults are injected in MSB, MSB-1, and fixed bit positions 25, 20, 10, and 5 of a single adder output. The accuracies for AlexNet, BERT, LLaMA-3.2-1B, and Qwen-0.6B-Base due to these faults are recorded as seen in Fig.~\ref{fig:cim_adders}(a) through (d), respectively. As mantissa sums propagate, the position of MSB drifts upward (bit 26 at level 1, bit 32 by level 7), so the significance of fixed-bit positions and the effect due to faults in them weaken mid-pipeline. However, they intensify after the global-alignment stage, which right-shifts mantissas. This is observed for bit 25, where the accuracies of AlexNet and BERT fall to $7.81\%$ and $45.63\%$ at level 1, rebound to $51.25\%$ and $88.44\%$ at level 4, then plunge to $5.94\%$ and $24.69\%$ at level 5 after global alignment. LLaMA-3.2-1B remains near $0\%$ across all levels, an observation due to its model depth. MSB and MSB-1 faults drive AlexNet and BERT to near-random accuracy and LLaMA-3.2-1B to $0\%$, whereas bits 20, 10, 5 are largely benign. Qwen-0.6B-Base matches LLaMA-3.2-1B for bit 25 ($0\%$ at every level) but shows a unique pattern for bit 20 due to the global alignment in the adder tree. The model maintains baseline accuracy through levels 1–4. The model accuracy drops to $10\%$ at level 5, going up to $33\%$ at level 7.

\begin{tcolorbox}[colback=green!5!white, colframe=green!75!black] \textbf{Observation 5:} Bit-flip faults in MSB and MSB–1 in the adder tree outputs are catastrophically amplified through successive accumulation levels and the global-alignment right shift, making the adder stage a critical hardware reliability bottleneck. Faults in bits between MSB and LSB (\textit{e.g.}, bit 25) exhibit increasing resilience till the global alignment stage, but become sensitive afterward, further exposing adder sensitivity.
\end{tcolorbox}

\subsubsection{\textbf{Global Alignment Faults}}
\label{subsubsec:global_align}

Fault injection at the global alignment target (1) offset values based on added groupwise exponents from pre-alignment; and (2) the added exponents themselves obtained from the pre-alignment stage. 

\begin{figure}[t]
    \centering
    \captionsetup{justification=centering}
    \includegraphics[width=0.95\linewidth]{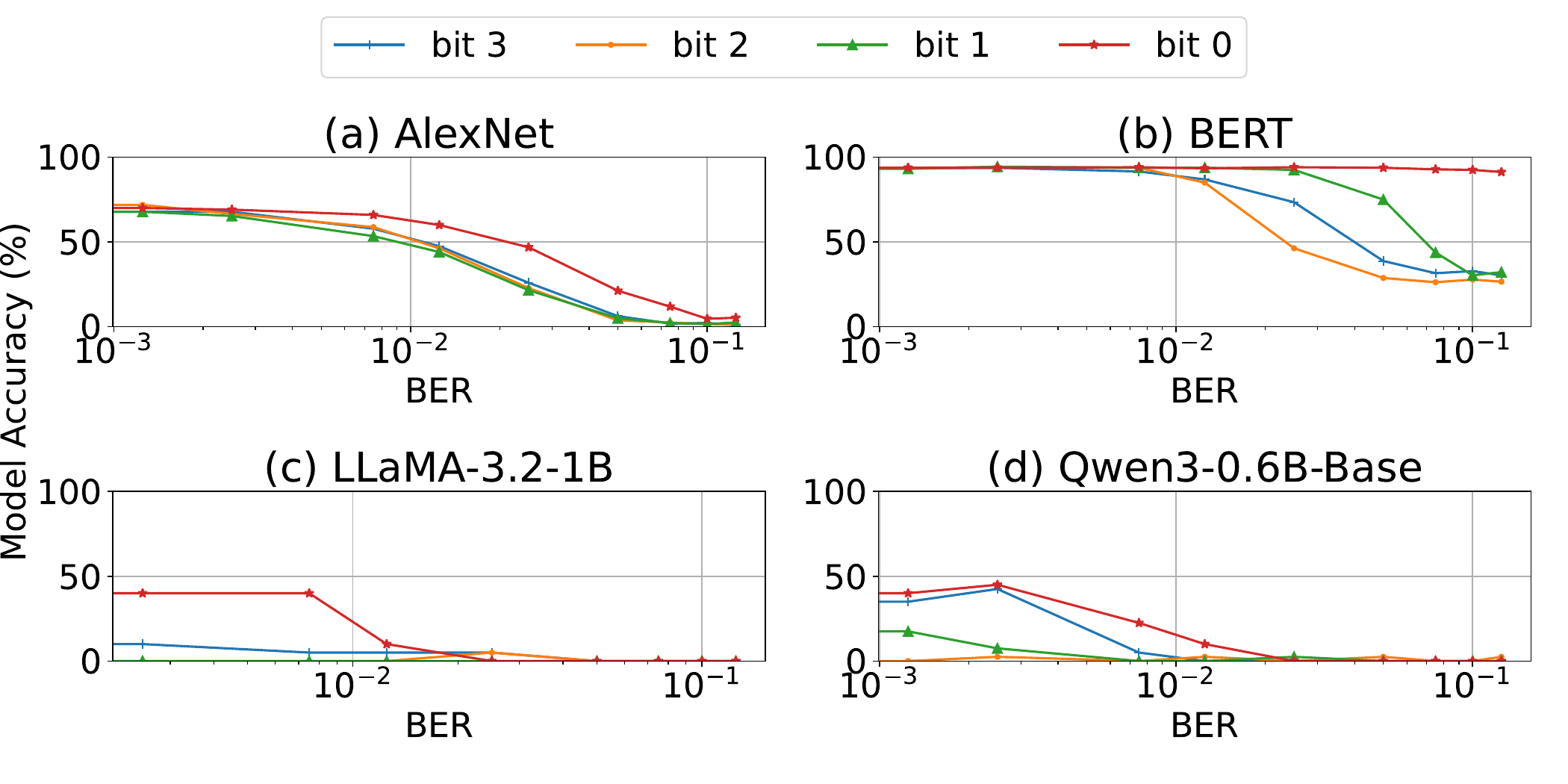}
    \caption{Accuracy after fault injection into offsets at Global Alignment.}
    \label{fig:cim_global_alignment_offsets}
\end{figure}

\paragraph{\textbf{Faults in Offsets}}\noindent To assess the effect of faults during offset computation in the global alignment stage, bit-flip faults are injected in the global alignment offsets encoded as 4-bit unsigned integers ($N_B=4$). Figs.~\ref{fig:cim_global_alignment_offsets}(a–d) summarize the effect of faults in offsets for AlexNet, BERT, LLaMA-3.2-1B, and Qwen3-0.6B-Base, respectively. For offset faults, low BER yields few misalignments and minor precision loss. An increase in BER translates to increased misalignments, leading to a greater accuracy drop. This is observed through faults in offset bits 3, 2, and 1 that reduce AlexNet accuracy to $6.25\%$, $3.75\%$, and $5\%$, respectively, at a BER of $5\times10^{-2}$. For the same BER, the accuracy of BERT drops to $28.75\%$ for bit 2 faults and $75\%$ for bit 1 faults. LLaMA-3.2-1B's accuracy falls to $10\%$ at $1.25\times10^{-2}$ BER in bit 1, and the same for Qwen3-0.6B-Base drops to $0\%$ at $1.25\times10^{-3}$ BER in bit 2. 

\begin{figure}[t]
    \centering
    \captionsetup{justification=centering}
    \includegraphics[width=0.95\linewidth]{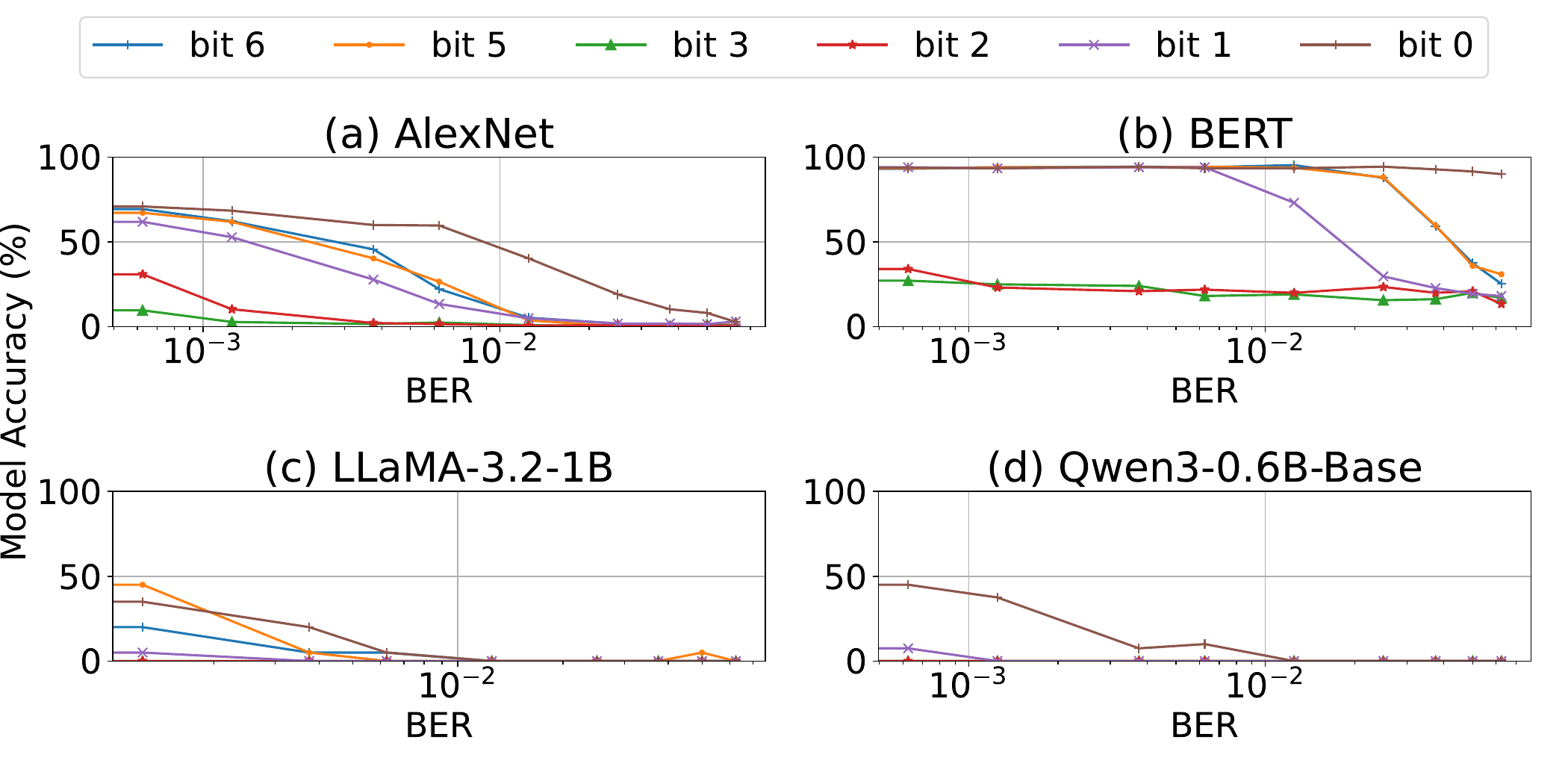}
    \caption{Accuracy after fault injection into exponents at Global Alignment.}
    \label{fig:cim_global_alignment_exp}
\end{figure}

\paragraph{\textbf{Faults in Exponents}}\noindent We inject bit-flip faults in the exponents at the global alignment stage to encoded as 8-bit unsigned integers ($N_B=8$). Figs.~\ref{fig:cim_global_alignment_exp}(a–d) illustrate the effect of faults in exponents for AlexNet, BERT, LLaMA-3.2-1B, and Qwen3-0.6B-Base, respectively. At $6.25\times10^{-4}$ BER, LLaMA-3.2-1B and Qwen3-0.6B-Base accuracies collapse to $0\%$, while AlexNet and BERT approach random performance ($9.69\%$ and $27.19\%$, respectively) for faults in exponent bit 3. Faults in exponent bits 2 and 3 skew the global maximum exponent upward (\textit{e.g.}, to values 125–127), inducing excessive right shifts and severe mantissa attenuation; bit 5 flips produce a broader exponent spread (values in $116–127$), yielding mixed shifts that partially cancel. MSB faults may generate NaNs, while LSB flips remain largely benign.

\begin{tcolorbox}[colback=green!5!white, colframe=green!75!black]
\textbf{Observation 6:} Bit-flips between MSB and LSB of the exponent (bits 2 and 3) in the global alignment directly corrupt its shift-control logic, causing column-wide mantissa mis-scaling. Faults in exponents exhibit higher sensitivity compared to offsets. 
\end{tcolorbox}

\subsubsection{\textbf{Normalization Faults}}
\label{subsubsec:normalization}

Single-bit faults are injected into the BFLOAT16 fields of normalized partial sums (sign, exponent, or mantissa) to simulate datapath faults in normalization. The results for AlexNet, BERT, LLaMA-3.2-1B, and Qwen3-0.6B-Base show that flipping bits in positions 15–8 leads to significant accuracy drops, such that model accuracy drops to $2.19\%$ (AlexNet), and $0\%$ (LLaMA-3.2-1B and Qwen3-0.6B-Base) at a BER of $3.125 \times 10^{-4}$. Bit-7 flips are less severe, with model accuracy dropping only to $47.81\%$, $33.5\%$, and $30\%$, respectively. BERT is more resilient to sign faults but still suffers from exponent flips. Mantissa faults also vary in impact, with bit 6 corruption at a BER of $3.125 \times 10^{-2}$ reducing accuracy to $37.5\%$ (AlexNet), $29.06\%$ (LLaMA-3.2-1B), and $0\%$ (Qwen3-0.6B-Base). Faults near LSB (such as in bits 2 and 3) show a notable accuracy drop only in BERT.

\begin{figure}[t]
    \centering
    \captionsetup{justification=centering}
    \includegraphics[width=0.95\linewidth]{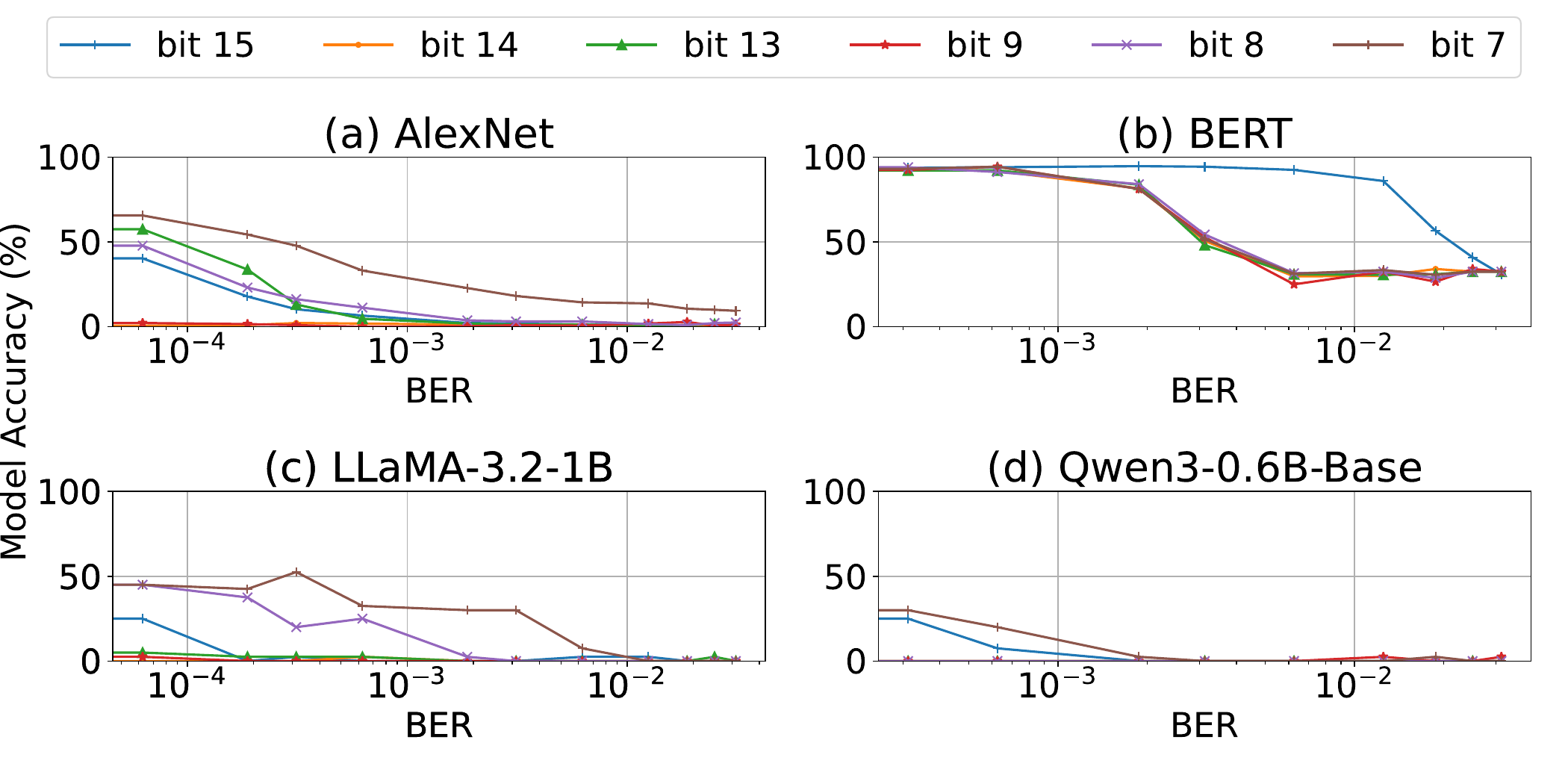}
    \caption{Accuracy due to faults in sign and exponent bits at normalization.}
    \label{fig:cim_normalization_exp}
\end{figure}

\begin{figure}[t]
    \centering
    \captionsetup{justification=centering}
    \includegraphics[width=0.95\linewidth]{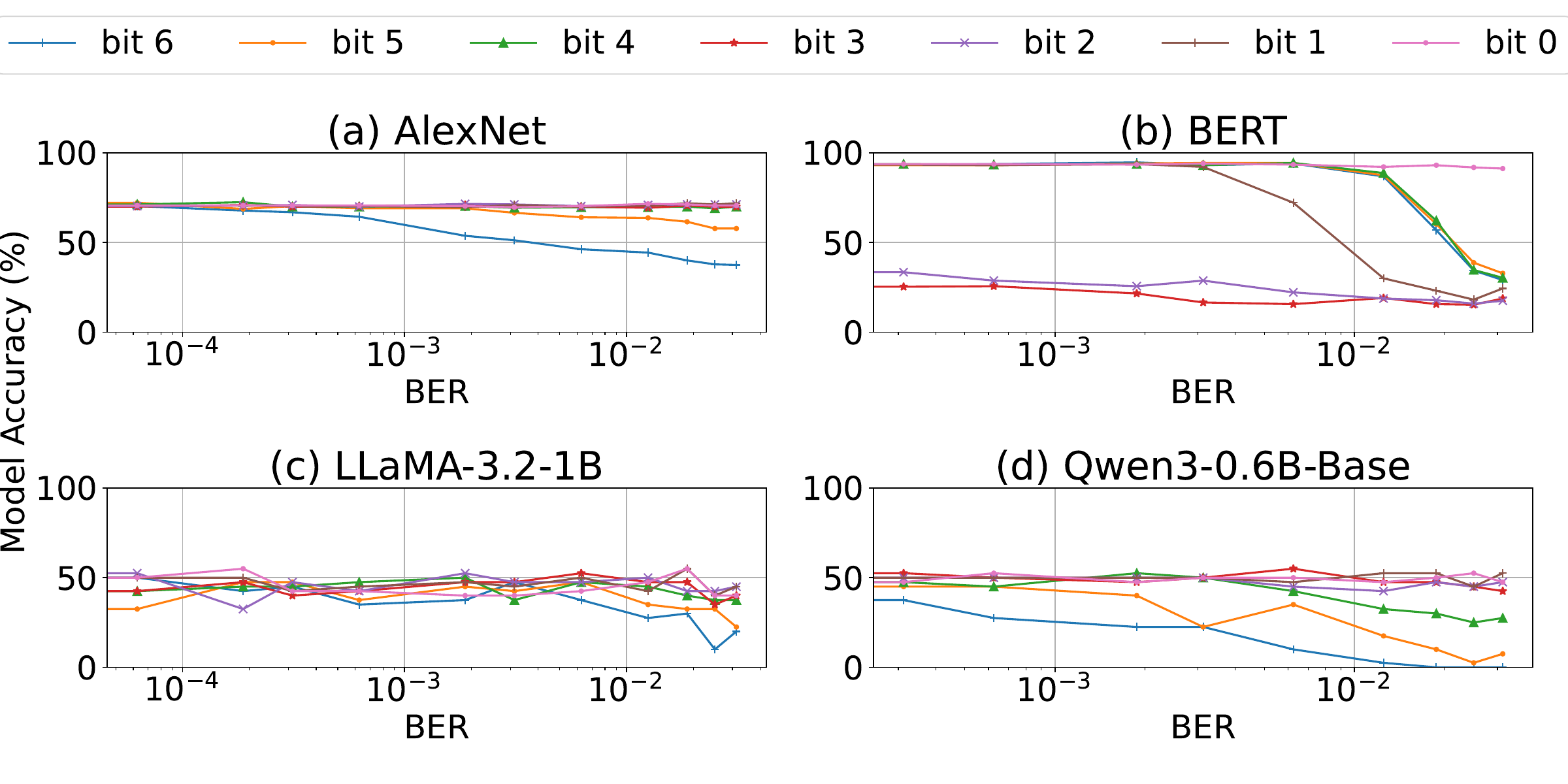}
    \caption{Accuracy due to faults in mantissa bits at the normalization stage.}
    \label{fig:cim_normalization_mant}
\end{figure}

\begin{tcolorbox}[colback=green!5!white, colframe=green!75!black] \textbf{Observation 7:} Normalization is the most fragile stage in FP-CiM. Faults in sign and exponent bits can severely corrupt outputs, often yielding NaNs. It is also observed that faults in different mantissa bits do not follow a general trend and are model dependent.
\end{tcolorbox}

\begin{table*}[htbp]
\centering
\caption{Summary of relative vulnerability of computation stages in the FP CiM pipeline.}
\begin{tabular}{|l|l|l|l|}
\hline
\multicolumn{1}{|c|}{\textbf{Stage}} & \multicolumn{1}{c|}{\textbf{Vulnerability Rank}} & \multicolumn{1}{c|}{\textbf{Primary Fault Impact}}                                                                                       & \multicolumn{1}{c|}{\textbf{Key Observations}}                                                                                                                      \\ \hline \hline
Normalization                        & Highest (1)                                      & \begin{tabular}[c]{@{}l@{}}Sign and Exponents cause \\ NaNs or polarity reversal\\ \textbf{(Observation 7)}\end{tabular}                         & \begin{tabular}[c]{@{}l@{}}Even minimal percentage of faulty computations \\ lead to collapse; BERT shows higher mantissa \\ sensitivity than other models.\end{tabular}                   \\ \hline
Global Alignment                     & High (2)                                         & \begin{tabular}[c]{@{}l@{}}Exponent faults introduce \\ misalignment and scaling errors\\ \textbf{(Observation 6)}\end{tabular}            & \begin{tabular}[c]{@{}l@{}}Bits 2 and 3 are most disruptive due to shift \\ in exponent ranges; offsets also disruptive \\ but less than exponents\end{tabular}     \\ \hline
Adder Tree                           & Significant (3)                                  & \begin{tabular}[c]{@{}l@{}}Faults in MSB/MSB-1 amplify \\ over adder levels\\ \textbf{(Observation 5)}\end{tabular}                               & \begin{tabular}[c]{@{}l@{}}Bit 25 shows non-monotonic behavior due to \\ global alignment shift; highly sensitive \\ to single-bit faults\end{tabular}              \\ \hline
Multipliers                          & Moderate (4)                                     & \begin{tabular}[c]{@{}l@{}}Faults in output MSBs cause \\ severe degradation\\ \textbf{(Observation 4)}\end{tabular}                              & \begin{tabular}[c]{@{}l@{}}LLaMA-3.2-1B drops to 0\% accuracy when \\ 0.1\% computations have MSB faults; \\ LSBs mostly benign\end{tabular}                                      \\ \hline
Memory Cells                         & Moderate (5)                                     & \begin{tabular}[c]{@{}l@{}}MSB faults in mantissas \\ propagate to all products\\ \textbf{(Observation 3)}\end{tabular}                           & \begin{tabular}[c]{@{}l@{}}Needs higher percentages of faulty computations  \\(\textgreater{}10\%) to show degradation; deeper models \\ accumulate small errors more rapidly\end{tabular} \\ \hline
Input Pre-Alignment                  & Low (6)                                          & \begin{tabular}[c]{@{}l@{}}Exponent/offset faults misalign \\ mantissas before memory \\ programming \textbf{(Observation 1)}\end{tabular} & \begin{tabular}[c]{@{}l@{}}Input exponent faults more damaging than \\ offsets; propagate across all output channels\end{tabular}                                \\ \hline
Weight Pre-Alignment                 & Lowest (7)                                       & \begin{tabular}[c]{@{}l@{}}Local weight mantissa misalignment \\ affects particular output channels\\ \textbf{(Observation 2)}\end{tabular} & \begin{tabular}[c]{@{}l@{}}Weight exponent faults worse than offsets; \\ limited propagation reduces impact\end{tabular}                                          \\ \hline
\end{tabular}
\label{tab:stage_summary}
\end{table*}

\subsubsection{\textbf{Effect of Quantization}} \label{subsubsec:quant}

\begin{figure}[t]
    \centering
    \captionsetup{justification=centering}

    \includegraphics[width=0.95\linewidth]{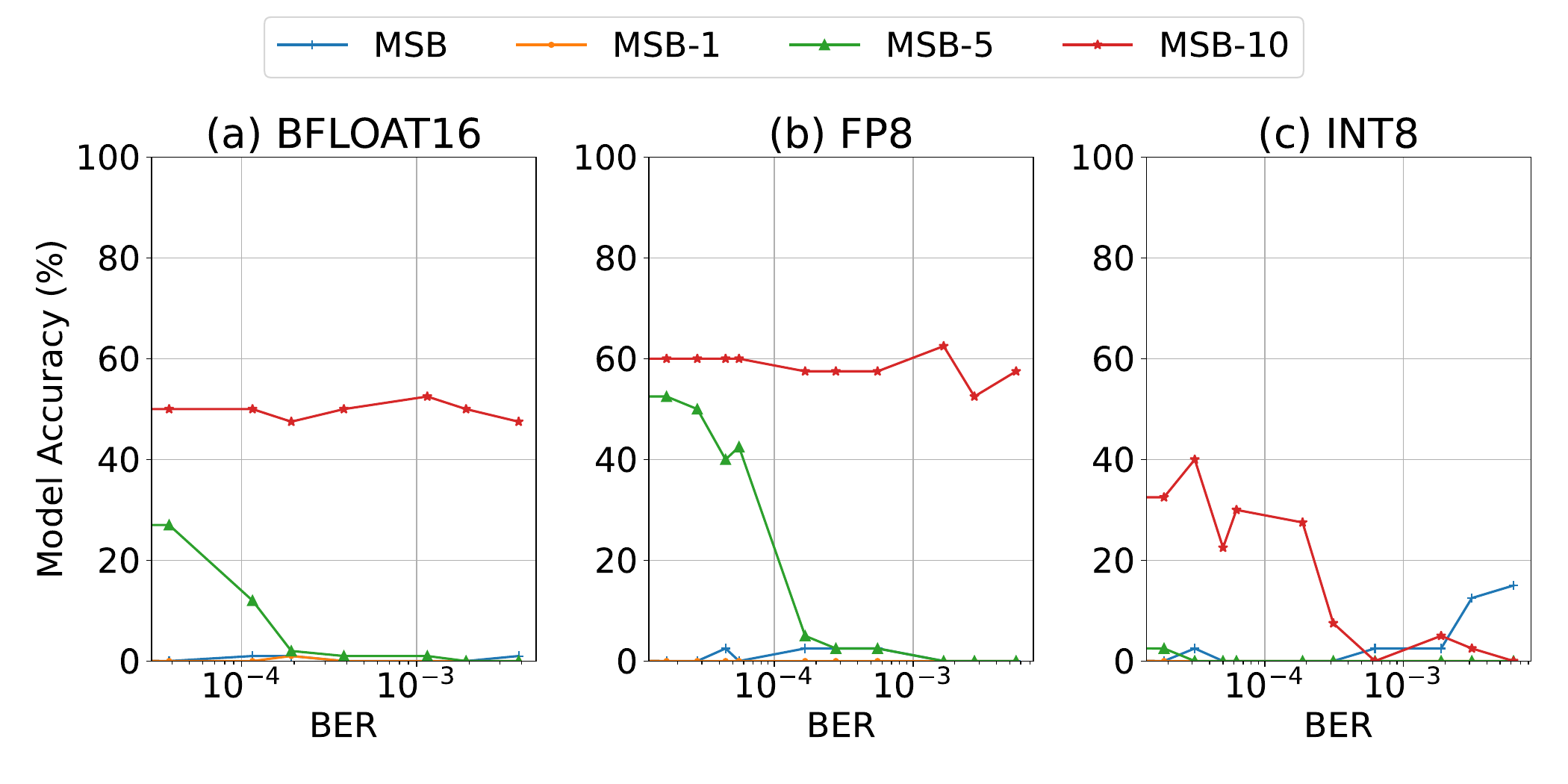}
    \caption{Multiplier faults in Qwen3-0.6B-Base for BFloat16, INT8 and FP8.}
    \label{fig:quant}
\end{figure}

To assess the impact of post-training quantization on fault sensitivity, we performed multiplier fault injections on Qwen3-0.6B-Base, adjusting the FP-CiM pipeline for FP8 (4-bit exponent and 3-bit mantissa) and INT8 (no alignment stages). The multiplier outputs are 26 bits for BFLOAT16, 18 bits for FP8, and 16 bits for INT8. Faults are injected at various bit positions, with accuracies shown in Fig.~\ref{fig:quant}(a–c). Quantization to a different FP datatype provided no resilience benefits. BFLOAT16 and FP8 maintained accuracy around the baseline with MSB–10 faults, with progressive accuracy degradation for faults in MSB-5 with increasing BER. Furthermore, INT8 accuracy dropped to $30\%$ at a low BER of $6.25\times10^{-5}$ and to $0\%$ at BER of $6.25\times10^{-4}$. Corrupting just $0.03\%$ of multiplier outputs led to zero accuracy across all datatypes for MSB faults. This indicates that INT8 is highly vulnerable to faults, whereas BFLOAT16 and FP8 are more stable against faults near LSBs. 

\begin{tcolorbox}[colback=green!5!white, colframe=green!75!black] \textbf{Observation 8:} Faults occurring near the LSB in INT8 computation exhibits increased sensitivity as compared to BFLOAT16 and FP8 computations. Thus, BFLOAT16 and FP8 computations are more resilient compared to INT8. 
\end{tcolorbox}

\subsection{Summary of Observations}
\label{subsec:observations}

\subsubsection{\textbf{Critical Component Analysis}} 
\label{subsubsec:critical_analysis}
Table~\ref{tab:stage_summary} reveals a fault‐sensitivity hierarchy of different FP-CiM computation stages based on our fault injection experiments. For each FP-CiM computation stage, we provide a vulnerability rank from Highest (denoted by 1) to Lowest (7). The table also shows the primary fault impact for each computation stage and key observations from the fault injection experiments. Comparing the accuracy degradation trends for each stage with respect to BER, we find that the normalization stage is the most sensitive, followed by global alignment, then adder tree, multipliers, memory, and then pre‐alignment. This mirrors the inverse order of computations in the FP-CiM. Faults in the later stages of the FP-CiM computation pipeline occur near to the output, minimizing correction opportunities. Stages with similar accuracy impact for a given BER are ordered by their positioning in the compute pipeline, since there is an opportunity for the fault effects to be masked by later stages. For example, a single fault in either the adder tree or global alignment can reduce accuracy to $0\%$. However, global alignment is deemed more sensitive due to its later placement in the compute pipeline.

\subsubsection{\textbf{Influence of Architecture and Models on Fault Effects}}
\label{subsubsec:effect_model_arch}
In this section, we discuss the effect that FP-CiM architecture would have on fault resilience, and how different model architectures can influence these effects.
\paragraph{\textbf{Architecture‐Driven Bit Sensitivity}} 
Within a computation stage, the significance of a bit sets the general trend that the model accuracy will follow when a fault occurs in that bit. For example, since the MSB has the highest bit significance, the accuracy degradation will be the most when a fault occurs in the MSB. Sign or exponent bit-flips in \emph{normalization} or \emph{global alignment} can incur catastrophic failures, either by inverting the sign or triggering runtime errors for MSB faults in exponents, regardless of the model or dataset. Faults in less significant mantissa bits produce more graded accuracy drops in proportion to bit significance. However, in global alignment (Section~\ref{subsubsec:global_align}), exponent bits~2 and~3 induce larger downstream mis‐scaling than the MSB, overturning the usual MSB‑dominance assumption.

\paragraph{\textbf{Model‐Driven Error Amplification}} 
A model’s depth, topology, and activation sparsity can impact the hardware fault effects. Sparse‐activation models (\textit{e.g.}\ BERT) suffer accuracy loss from mantissa faults in every bit except the LSB, whereas dense models (AlexNet, LLaMA‑3.2‑1B) only fail when faults occur in more significant mantissa bits (bits 6 or 5). Model depth further amplifies fault effects. Injecting MSB faults into $0.1\%$ of mantissas across LLaMA‑3.2‑1B’s decoder layers raises median per‑decoder block error. In contrast, the much shallower AlexNet remains robust until $3\%$ of the MSBs flip. This error in the decoder block outputs of LLaMA-3.2-1B is quantified by the median of relative absolute error between the outputs of the decoder blocks from the fault-injected inference and fault-free inference. This median per-decoder block error increases from $431.25\%$ to $721.88\%$, $943.75\%$, and $975\%$ in layers 1, 2, 4, and 6, respectively. A deviation of $140.62\%$ in final logits drives accuracy to $0\%$. Thus, while bit‐significance sets the \emph{potential} for error, a network’s structure and data patterns determine the \emph{realized} accuracy degradation. 
\section{Fault Resilient Design - \fram{}}
\label{sec:design}

Our fault injection experiments in Section~\ref{subsec:exp_analysis} provide insights into the resilience of FP-CiM accelerators. In this section, we explore different design characteristics and propose, for the first time, an informed guideline for designing a fault-tolerant digital FP CiM accelerator, \textbf{\fram{}}.

\subsection{Choosing Alignment Paradigm}
\label{subsec:design_alignment}

\begin{figure}
    \centering
    \captionsetup{justification=centering}
    \includegraphics[width=0.95\linewidth]{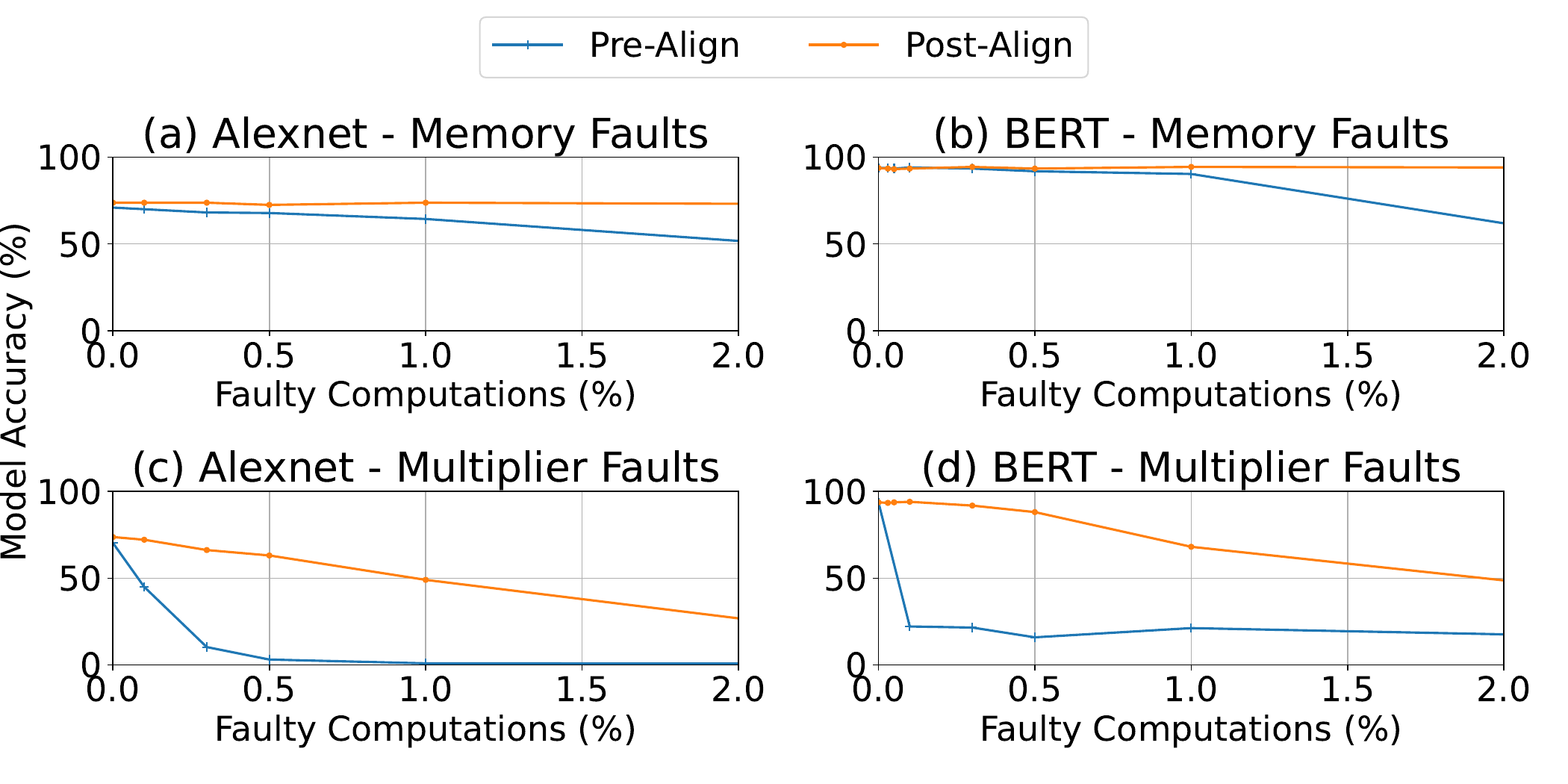}
    \caption{Fault resilience due to post-alignment design.}
    \label{fig:decision_alignment}
\end{figure}

In this section, we examine the alignment stage's positioning in the FP-CiM pipeline as a design decision for \textbf{\fram{}}. In pre-alignment designs, mantissas shift before arithmetic operations. This can worsen the impact of faults in CiM memory cells, multiplier, or adder outputs due to precision loss. In contrast, post-alignment shifts occur after multiplication, reducing the effect of MSB faults. For instance, if a mantissa product must shift by 5 bits, an MSB fault in the multiplier output introduces an error in the order of \(2^{-2}\) in a post-alignment design, compared to \(2^{3}\) in a pre-alignment design.

Figs.~\ref{fig:decision_alignment}(a) and (b) compare AlexNet and BERT under programmed-weight faults, respectively, while Figs.~\ref{fig:decision_alignment}(c) and (d) do the same with multiplier output faults. With faults in the MSBs of $0.5\%$ of the programmed weight mantissas, post-alignment preserves both models’ baseline accuracy. When faults are injected into the MSBs of $0.5\%$ of the multiplier outputs, AlexNet accuracy drops to $63.125\%$ for post-alignment design versus $1.25\%$ in pre-alignment design. BERT accuracy drops to $93.47\%$ for post-alignment design versus $46.25\%$ in a pre-alignment design for the same percentage of faulty multipliers. These results demonstrate that adopting a post-alignment design markedly increases fault resilience.

\begin{tcolorbox}[colback=cyan!5!white, colframe=cyan!75!black] \textbf{Design Decision 1:} For \textbf{\fram{}}, we adopt a post-alignment architecture (mantissa alignment after multiplication/addition rather than before) to reduce the effect MSB faults and improve fault resilience.
\end{tcolorbox}

\subsection{Choosing FP-CiM Stencil}
\label{subsec:design_alignment}

\begin{figure}
    \centering
    \captionsetup{justification=centering}
    \includegraphics[width=0.95\linewidth]{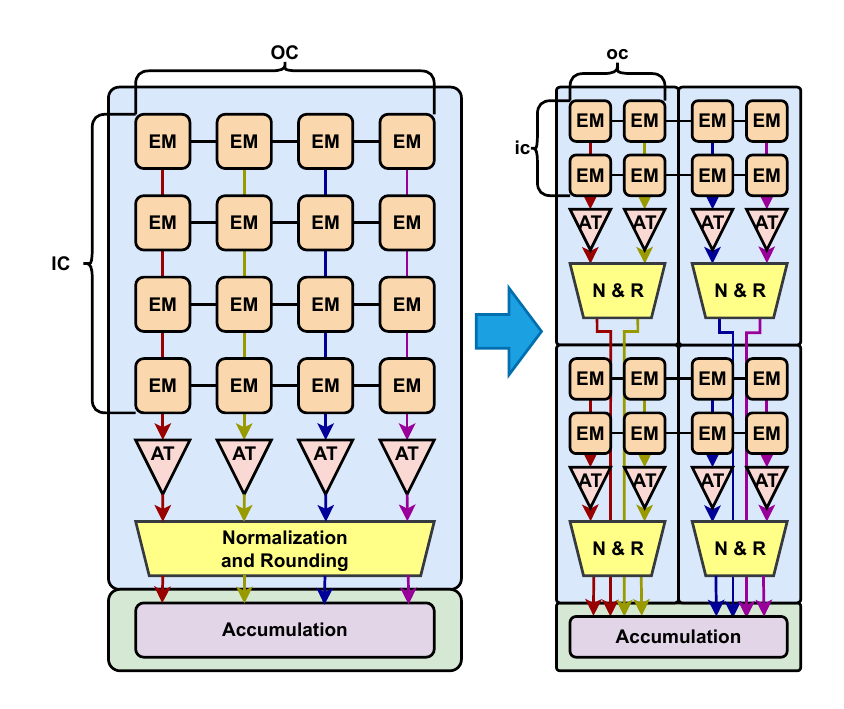}
    \caption{\textbf{ic $\times$ oc $\times$ H $\times$ W} stencil opposed to \textbf{IC $\times$ OC} stencil.}
    \label{fig:icochw_stencil}
\end{figure}

\begin{figure}
    \centering
    \captionsetup{justification=centering}
    \includegraphics[width=0.95\linewidth]{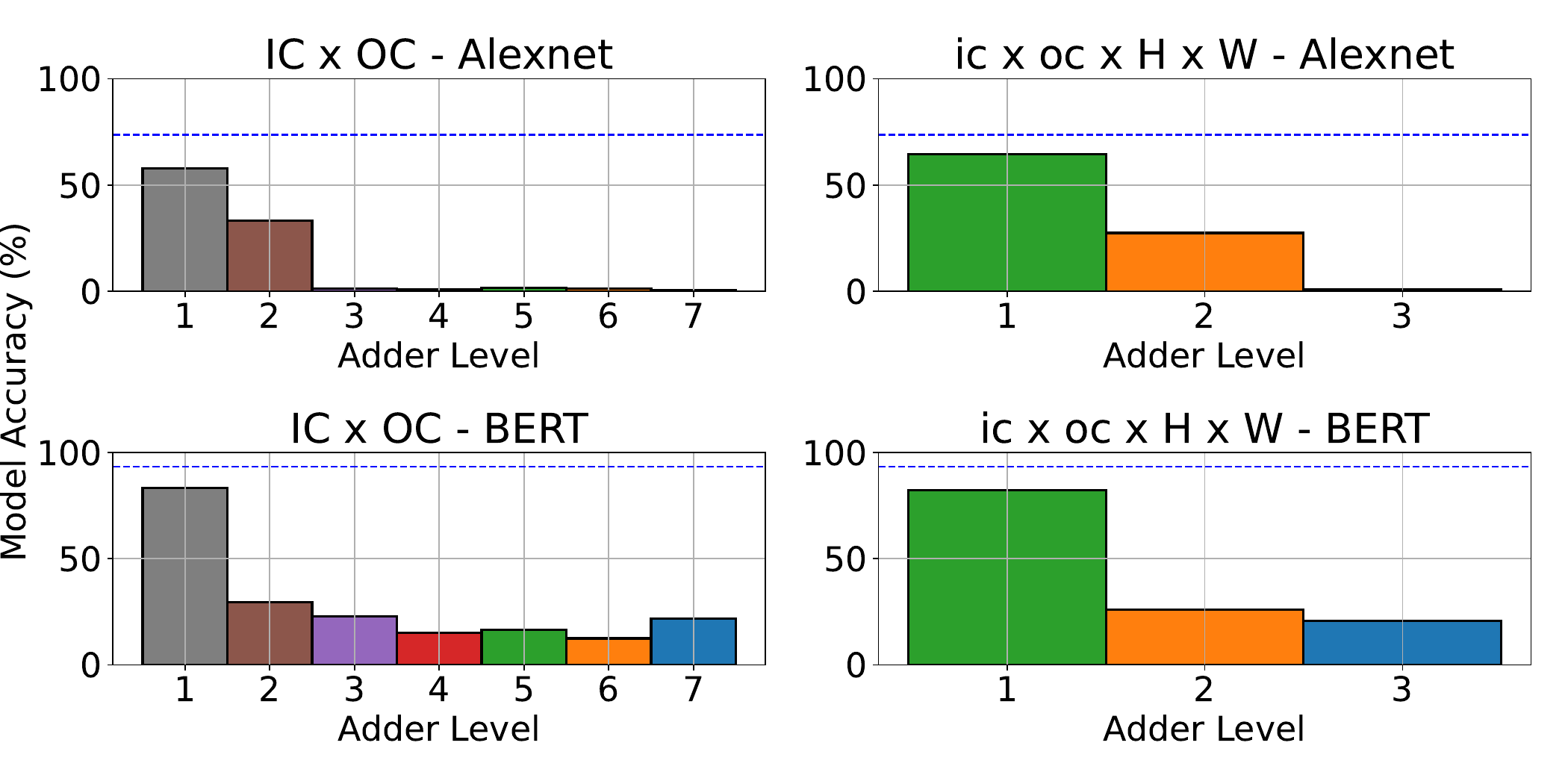}
    \caption{Fault resilience due to changing stencils.}
    \label{fig:decision_stencil}
\end{figure}

In this section, we assess changing the FP-CiM design stencil from the current \textbf{IC $\times$ OC} to \textbf{ic $\times$ oc $\times$ H $\times$ W} for improving fault resilience in post-alignment design for \textbf{\fram{}}. While post‐alignment design masks faults in the programmed weights and multiplier outputs, it does not mitigate faults in the adder outputs. This stems from our use of an \textbf{IC $\times$ OC} stencil, where we have a single adder tree of $\log_2{\textbf{IC}}$ levels with a global alignment stage followed by a normalization stage for a crossbar column as shown in the left side of Fig.~\ref{fig:icochw_stencil}. As the mantissa bit‐width increases at each adder level, an MSB fault early in the tree propagates through more levels, compounding its effect on model accuracy.

To reduce adder‐tree depth, we introduce an \textbf{ic $\times$ oc $\times$ H $\times$ W} stencil for \textbf{\fram{}} that partitions the \textbf{IC $\times$ OC} crossbar into smaller \textbf{ic $\times$ oc} crossbars such that:

\[
\textbf{IC} = \textbf{ic} \times \textbf{H}; \textbf{OC} = \textbf{oc} \times \textbf{W}
\]

Thus, due to the partitioning, there will be \textbf{H $\times$ W} smaller crossbars of \textbf{ic $\times$ oc} memory cells. As \textbf{ic} is smaller compared to \textbf{IC}, the adder trees will be smaller, having $\log_2{\textbf{ic}}$ levels. Each of the \textbf{H $\times$ W} partitions now would have its own normalization and rounding stage, as shown in the right side of
Fig.~\ref{fig:icochw_stencil}. Due to the smaller adder trees, the errors due to faults in an adder have to percolate through fewer adder levels. This is validated through an experiment on AlexNet with CIFAR100 in Fig.~\ref{fig:decision_stencil} for \textbf{ic=}8, \textbf{oc=}4, \textbf{H=16} and \textbf{W=8}. In each pair of plots, the top two depict AlexNet results and the bottom two depict BERT; left panes correspond to \textbf{IC $\times$ OC}, and right panes to \textbf{ic $\times$ oc $\times$ H $\times$ W}. It shows that at the lowest adder level (level 1), MSB faults have a markedly smaller impact under the \textbf{ic $\times$ oc $\times$ H $\times$ W} stencil than under the \textbf{IC $\times$ OC} stencil with \textbf{IC=}128 and \textbf{OC=}32. Specifically, AlexNet inference accuracy rises from $57.81\%$ to $64.65\%$, and BERT‐Base accuracy from $81.75\%$ to $84\%$. Thus, changing the FP-CiM design stencil from \textbf{ic $\times$ oc $\times$ H $\times$ W} markedly increases the fault resilience.

\begin{tcolorbox}[colback=cyan!5!white, colframe=cyan!75!black] \textbf{Design Decision 2:} We consider a \textbf{ic $\times$ oc $\times$ H $\times$ W} stencil for \textbf{\fram{}} over an \textbf{IC $\times$ OC} stencil for a more fault tolerant design.\end{tcolorbox}

\subsection{Choosing Configuration of FP-CiM Stencil}
\label{subsec:design_stencil_ablation}

\begin{figure}
    \centering
    \captionsetup{justification=centering}
    \includegraphics[width=0.95\linewidth]{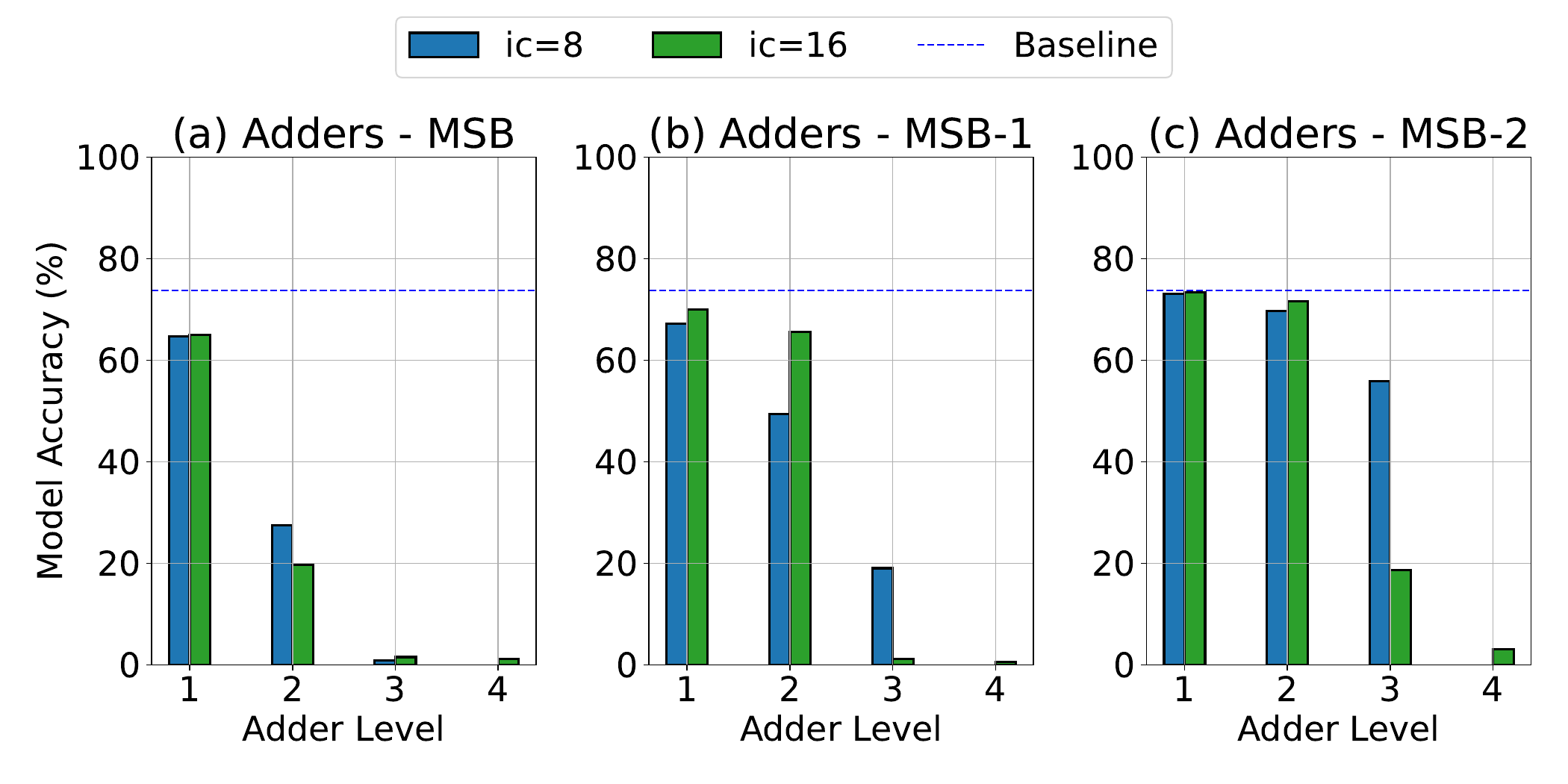}
    \caption{Comparison of different ablations of the \textbf{ic $\times$ oc $\times$ H $\times$ W}.}
    \label{fig:design_icochw_ablation}
\end{figure}

In this section, we investigate different configurations of the \textbf{ic $\times$ oc $\times$ H $\times$ W} stencil for \textbf{\fram{}} that can provide an improved fault resilience. Varying \textbf{ic} and \textbf{oc} introduces microarchitectural changes in the \textbf{\fram{}} that can affect how the fault effects propagate through an FP-CiM pipeline. Reducing \textbf{ic} for a constant total number of MAC units makes the adder trees smaller. Thus, the effect of faults occurring in the adder tree would be percolating down smaller number of adder levels, reducing the accuracy degradation. For a constant total number of MAC units, varying \textbf{oc} for constant product \textbf{ic $\times$ H} has no effect. This is because the configuration of \textbf{oc} and \textbf{W} have no effect on adder tree depth. Thus, for \textbf{\fram{}}, we vary the value of \textbf{ic}. Fig.~\ref{fig:design_icochw_ablation} plots AlexNet accuracy on CIFAR100 for adder faults at different adder-tree levels for total MAC units of 4096, comparing configuration $8 \times 4 \times 16 \times 8$ and $16 \times 4 \times 8 \times 8$ of the \textbf{ic $\times$ oc $\times$ H $\times$ W} stencil. We observe that smaller \textbf{ic} reduces accuracy degradation by limiting adder levels. Fault injection in a single adder shows that for MSB faults, the model accuracy for \textbf{ic}=8 and \textbf{ic}=16 are similar when the fault occurs in adder level 1. The accuracy improves for \textbf{ic}=8 for a fault at adder level 2 for MSB faults. MSB-1 and MSB-2 faults also lead to smaller accuracy degradation for \textbf{ic}=8 for a fault in adder level 3. Therefore, we choose the configuration $8 \times 4 \times 16 \times 8$ for our \textbf{\fram{}}. 

\begin{tcolorbox}[colback=cyan!5!white, colframe=cyan!75!black] \textbf{Design Decision 3:} The configuration of the \textbf{ic $\times$ oc $\times$ H $\times$ W} affects fault resilience, with smaller \textbf{ic} leading to greater fault resilience due to smaller adder trees.\end{tcolorbox}

\subsection{Choosing Global Alignment}
\label{subsec:design_ga}

\begin{figure}
    \centering
    \captionsetup{justification=centering}
    \includegraphics[width=0.95\linewidth]{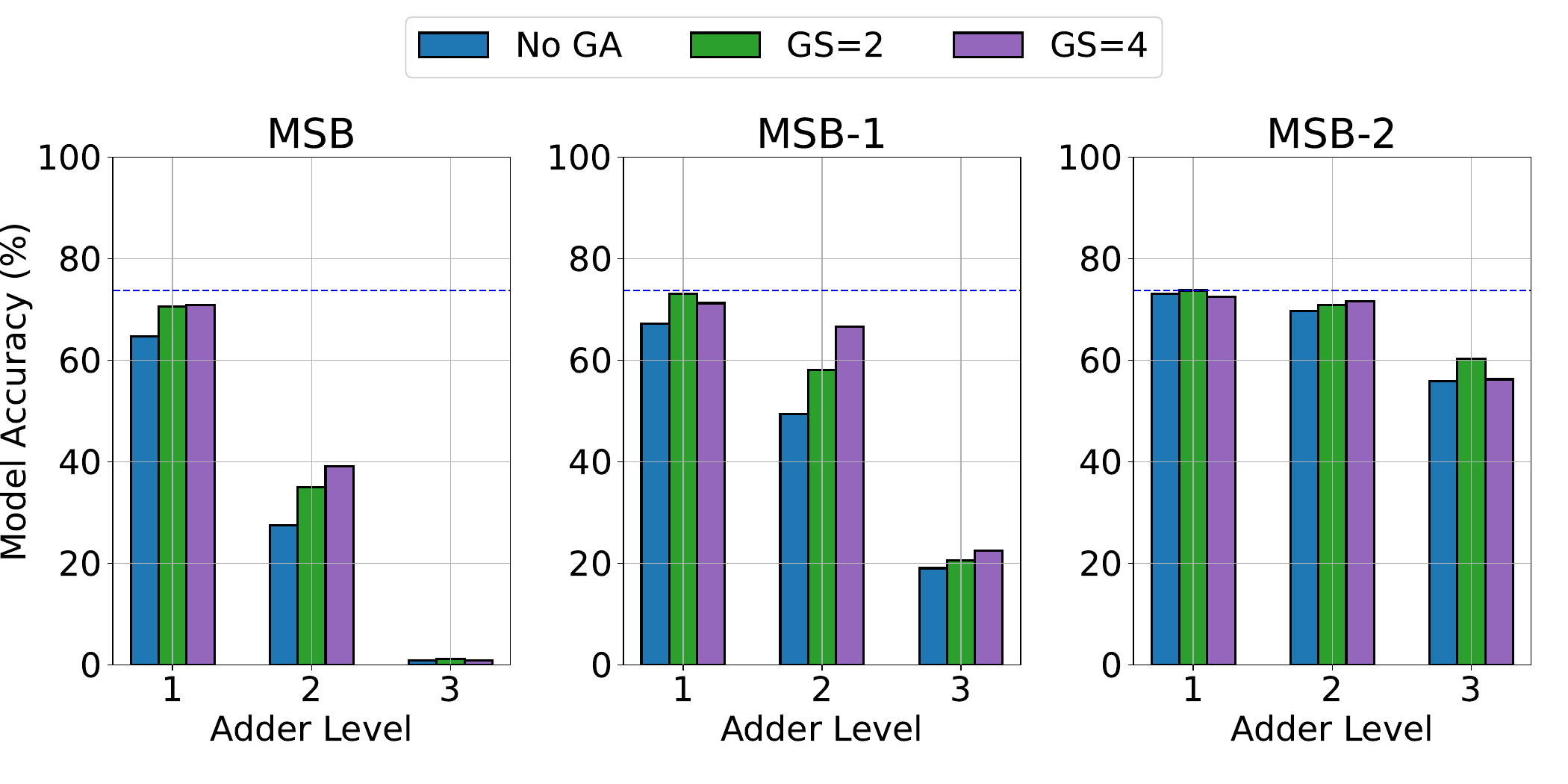}
    \caption{Comparing FP-CiM crossbar with global alignment.}
    \label{fig:design_ga}
\end{figure}

In this section, we investigate if inclusion of a global-alignment stage in \textbf{\fram{}} will aid in improving the fault resilience of the FP-CiM. In our assessments so far, we have observed that the positioning of the alignment stage has an effect on fault resilience. If the alignment stage is before all the arithmetic operations (pre-alignment), the fault resilience reduces as observed in Section~\ref{subsec:design_alignment}. When the alignment occurs after multiplication (post-alignment), we observe an improved resilience towards multiplier faults. Thus, alignment stages can effectively mask faults in the datapath when it is occuring after an arithmetic stage. Therefore, we propose having a two-staged alignment process in \textbf{\fram{}}, where we first align the multiplier outputs in groups (local alignment), and then the mantissa sums at an adder level determined by the local alignment grouping. We call the adder level alignment as global alignment. This two-staged alignment is in line with the design in~\cite{kaul2019optimized}. Fig.~\ref{fig:design_ga} shows AlexNet accuracy on CIFAR100 showing degradation due to faults in the adder tree for \textbf{ic $\times$ oc $\times$ H $\times$ W} stencil with configuration $8 \times 4 \times 16 \times 8$ and $16 \times 4 \times 8 \times 8$. The accuracies under adder faults are compared for a design with (1) no global alignment, (2) global alignment for local alignment with a group size of 2, and (3) global alignment for local alignment with a group size of 4. It is evidenced that local alignment followed by global alignment leads to a comparatively smaller drop in accuracy as compared to a design with no global alignment. Specifically, for MSB-1 fault occurring in a single adder at adder level 1 for a group size of 4 in local alignment, the accuracy drops only to $69.51\%$. Comparing this to the same fault for our pre-alignment design in Section~\ref{subsubsec:cim_adders} (bit 25 is MSB-1 at level 1 where accuracy is $7.81\%$), the accuracy drop from the baseline of $70.81\%$ is \textbf{49$\times$} less in this \textbf{ic $\times$ oc $\times$ H $\times$ W} design with local and global alignment. 

\begin{tcolorbox}[colback=cyan!5!white, colframe=cyan!75!black] \textbf{Design Decision 4:} We incorporate a global alignment stage within the adder tree in our \textbf{\fram{}} with a post-alignment design to enhance fault resilience.\end{tcolorbox}

\subsection{Designing a Fault Tolerant FP-CiM for 4096 MACs}
\label{subsec:design_summary}

\begin{table}[h]
    \centering
    \caption{Specifications for fault-tolerant FP-CiM design.}
    \label{tab:design_summary}
    \begin{tabular}{ll}
        \toprule
        \textbf{Design Decision} & \textbf{Specification} \\
        \midrule
        Total MACs & $4096$ \\
        Floating-Point Format & BFLOAT16 \\
        Mantissa Bits & 7 bits (padded to 12 bits) \\
        Mantissa Representation & 13 bits \\
        Alignment Paradigm    & Post-Alignment \\
        Stencil & \textbf{ic $\times$ oc $\times$ H $\times$ W} \\
        Stencil Configuration & $8 \times 4 \times 16 \times 8$ \\
        Adder Levels & 3 stages \\
        Global Alignment  & After two adder levels (group size of 4) \\
        \bottomrule
    \end{tabular}
\end{table}

Table~\ref{tab:design_summary} summarizes the design decisions taken with their specifications for a fault-tolerant FP-CiM with 4096 MACs using \textbf{\fram{}}. To design an FP-CiM with 4096 MACs, we first choose an \textbf{ic $\times$ oc $\times$ H $\times$ W} stencil. For this stencil, we choose \textbf{ic $\times$ H = IC} to be 128, and \textbf{oc $\times$ W = OC} to be 32 as discussed in Section~\ref{subsec:exp_setup}. After selecting the stencil, we choose our alignment stage to occur after CiM multiplication, thus opting for a post-alignment design. After this, to keep the adder trees small, we select \textbf{ic}=$8$, thus opting for the \textbf{ic $\times$ oc $\times$ H $\times$ W} configuration of $8 \times 4 \times 16 \times 8$. Furthermore, for the same configuration, we select and propose to have a local alignment stage after CiM multiplication with a group size of $4$ followed by global alignment in the adder tree in our post-alignment design. \textbf{\fram{}} designed using these design decisions shows a 49$\times$ improvement in accuracy drop for a single adder fault in AlexNet. 

\section{Conclusion}
\label{sec:conclusion}

This work presents the first in-depth fault resilience study of digital FP CiM accelerators, leveraging our custom fault injection framework, \textbf{FaultCiM}. Through computation stage-wise, bit-level analysis across various model and dataset combinations, %AlexNet, BERT, and LLaMA-3.2-1B, 
we uncover how fault impact varies with architectural stage, bit significance, and model-specific traits such as depth and sparsity. Our results identify normalization as the most fragile stage, where exponent and sign bit faults lead to catastrophic errors. Deeper models like LLaMA-3.2-1B are particularly vulnerable due to cumulative fault effects. %, even from mid-bit faults. 
Our fault resilience study further reveals that microarchitectural design decisions such as pre-alignment group size or alignment stage positioning can significantly influence fault resilience. We use this analysis to provide a set of design decisions that could be used to design a fault-resilient FP-CiM accelerator,  \textbf{\fram{}}. Using the design decisions, we propose \textbf{\fram{}} for 4096 MACs that can improve the accuracy drop from the baseline by 49$\times$ for a single adder fault compared to a naive pre-alignment design.

\section*{Acknowledgments}
This research is funded by National Science Foundation through NSF grant number 2537757.

\balance
\bibliographystyle{IEEEtran}
\bibliography{ref}

\end{document}